\documentclass{aa}  

\usepackage{graphicx}
\usepackage{txfonts}
\usepackage{lipsum}
\usepackage{subcaption}         
\usepackage{lscape}             
\usepackage{placeins}           
\usepackage{comment}
                                
\DeclareRobustCommand{\VAN}[3]{#2}
\let\VANthebibliography\thebibliography
\def\thebibliography{\DeclareRobustCommand{\VAN}[3]{##3}\VANthebibliography}

\usepackage{hyperref}
\usepackage{gensymb}

\usepackage{xcolor}
\definecolor{linkcolor}{rgb}{0,0,0.5}
\hypersetup{colorlinks=true, linkcolor=linkcolor, citecolor=linkcolor, filecolor=linkcolor, urlcolor=linkcolor}

\newcommand\MgII{\hbox{Mg$\,\rm \scriptstyle II$}~} 
\newcommand\HI{\hbox{H$\,\rm \scriptstyle I$}}

\newcommand\OI{\hbox{O$\,\rm \scriptstyle I$}~} 

\newcommand\Lya{Ly$\alpha$} 
\newcommand\Lyb{Ly$\beta$~}

\usepackage{xcolor}
\def\tomas#1{\textcolor{blue}{TS: #1}}

\usepackage{fancyhdr}

\fancypagestyle{plain}{
  \fancyhf{}
  \fancyfoot[C]{\thepage}
  
}

\usepackage[switch]{lineno}

\begin{document}


      
   \title{Forest without Trees is still Fruitful:}

   \titlerunning{Limits on the Neutral IGM Thermal State from the 21-cm Forest}

   \subtitle{Constraints on the thermal state of the neutral IGM at $z\approx5.6$ with the 21-cm forest power spectrum}

%

   \author{T. Šoltinský\inst{1,2,3,4}\fnmsep\thanks{Corresponding author: tomas.soltinsky@inaf.it}
        \and A. Chakraborty\inst{5,6}
        \and G. Kulkarni\inst{7}
        \and M. Viel\inst{1,2,3,4}
        \and C. M. Trott\inst{8,9,10}
        \and R. Sagar\inst{11}
        \and N. Thyagarajan\inst{12}
        \and J. S. Bolton\inst{13}
        \and B. Ciardi\inst{14}
        \and E. V. Ryan-Weber\inst{15}
        \and S. Maitra\inst{7}
        \and A. Datta\inst{11}
        \and N. Roy\inst{16}
        }

   \institute{INAF–Osservatorio Astronomico di Trieste, Via G.B. Tiepolo, 11, I-34143 Trieste, Italy
    \and INFN, Sezione di Trieste, Via Valerio 2, I-34127 Trieste, Italy
    \and SISSA, International School for Advanced Studies, Via Bonomea 265, 34136 Trieste, Italy
    \and IFPU, Institute for Fundamental Physics of the Universe, Via Beirut 2, I-34151 Trieste, Italy
    \and Department of Physics, McGill University, Montréal, QC, Canada
    \and Trottier Space Institute, McGill University, 3550 rue University, Montréal, QC H3A 2A7, Canada
    \and Tata Institute of Fundamental Research, Homi Bhabha Road, Mumbai 400005, India
    \and International Centre for Radio Astronomy Research, Curtin University, Bentley, WA 6102, Australia
    \and ARC Centre of Excellence for All Sky Astrophysics in 3 Dimensions (ASTRO-3D)
    \and Australia Telescope National Facility, CSIRO, Space and Astronomy, Bentley, WA, Australia
    \and Department of Astronomy, Astrophysics and Space Engineering, Indian Institute of Technology Indore, Indore 452020, India
    \and Space \& Astronomy, Commonwealth Scientific and Industrial Research Organisation (CSIRO), P. O. Box 1130, Bentley, WA 6102, Australia
    \and School of Physics and Astronomy, The University of Nottingham, University Park, Nottingham, NG7 2RD, UK
    \and Max-Planck-Institut für Astrophysik, Karl-Schwarzschild-Str. 1, D-85748 Garching b. München, Germany
    \and Centre for Astrophysics and Supercomputing, Swinburne University of Technology, P.O. Box 218, Hawthorn, VIC 3122, Australia
    \and Department of Physics, Indian Institute of Science, Bangalore 560012, India
    }

   \date{Received January XX, 20XX}

 
  \abstract
{Neutral regions of the intergalactic medium (IGM) during the Epoch of Reionization (EoR) are still largely unexplored due to the limitations of currently available probes. Owing to discoveries of numerous high-redshift radio-bright sources, the 21-cm forest, a series of absorption features imprinted by the neutral IGM in the spectra of such sources, now offers an attractive probe of the thermal and ionization state of the predominantly neutral IGM at $z\gtrsim5.5$.}
{We analyse archival upgraded Giant Metrewave Radio Telescope (uGMRT) observations of J352--15, the brightest known radio-loud quasar in the EoR ($z=5.82$), to measure the one-dimensional (1D) power spectrum of the 21-cm forest and constrain the thermal and ionization state of the neutral IGM.}
{We process $17.5\,\rm hr$ of Band-2 uGMRT observations covering $209.0$--$222.5\,\rm MHz$. Using the measured 1D power spectrum, we perform Bayesian parameter inference by comparing the observations with forward-modelled synthetic spectra generated from cosmological simulations spanning a wide range of ionization and X-ray pre-heating scenarios. This framework enables parameter inference even from a null detection. We also present an independent Murchison Widefield Array (MWA) measurement, although its lower sensitivity prevents competitive constraints.
}
{We achieve a sensitivity of $3.62\,\rm mJy\,beam^{-1}$ per $6.1\,\rm kHz$ channel. While we do not detect the 21-cm forest statistically, the null detection jointly constrains the mean neutral hydrogen fraction, $\langle x_{\rm HI}\rangle$, and the mean temperature of the neutral IGM, $\langle T_{\rm HI}\rangle$. At the $68\%$ credible level, our analysis disfavours cold and substantially neutral IGM models at $z\approx5.6$, including models with $\langle T_{\rm HI}\rangle \lesssim 27\,\rm K$ for $\langle x_{\rm HI}\rangle=0.1$. These limits probe regions of parameter space allowed by existing Ly$\alpha$ and 21-cm observations, suggesting substantial pre-heating of the neutral IGM above the adiabatic cooling floor.}
{The 21-cm forest has entered the regime of observationally informative statistics. Its 1D power spectrum provides an independent probe of the neutral IGM even in the absence of a detection.}



\twocolumn[
\begin{@twocolumnfalse}

\begin{center}
{\Large\bfseries
Forest without Trees is still Fruitful:\\
Constraints on the thermal state of the neutral IGM at $z\approx5.6$ with the 21-cm forest power spectrum
\par}

\vspace{0.5cm}

T. Šoltinský$^{1,2,3,4,*}$,
A. Chakraborty$^{5,6}$,
G. Kulkarni$^{7}$,
M. Viel$^{1,2,3,4}$,
C. M. Trott$^{8,9,10}$,
R. Sagar$^{11}$,
N. Thyagarajan$^{10}$
J. S. Bolton$^{12}$,
B. Ciardi$^{13}$,
E. V. Ryan-Weber$^{14}$,
S. Maitra$^{7}$,
A. Datta$^{11}$,
N. Roy$^{15}$

\vspace{0.4cm}
\end{center}

{\small
$^{1}$ INAF--Osservatorio Astronomico di Trieste, Via G.B. Tiepolo 11, I-34143 Trieste, Italy\\
$^{2}$ INFN, Sezione di Trieste, Via Valerio 2, I-34127 Trieste, Italy\\
$^{3}$ SISSA, International School for Advanced Studies, Via Bonomea 265, 34136 Trieste, Italy\\
$^{4}$ IFPU, Institute for Fundamental Physics of the Universe, Via Beirut 2, I-34151 Trieste, Italy\\
$^{5}$ Department of Physics, McGill University, Montréal, QC, Canada\\
$^{6}$ Trottier Space Institute, McGill University, 3550 rue University, Montréal, QC H3A 2A7, Canada\\
$^{7}$ Tata Institute of Fundamental Research, Homi Bhabha Road, Mumbai 400005, India\\
$^{8}$ International Centre for Radio Astronomy Research, Curtin University, Bentley, WA 6102, Australia\\
$^{9}$ ARC Centre of Excellence for All Sky Astrophysics in 3 Dimensions (ASTRO-3D), Australia\\
$^{10}$ Space \& Astronomy, Commonwealth Scientific and Industrial Research Organisation (CSIRO), P. O. Box 1130, Bentley, WA 6102, Australia\\
$^{11}$ Department of Astronomy, Astrophysics and Space Engineering, Indian Institute of Technology Indore, Indore 452020, India\\
$^{12}$ School of Physics and Astronomy, The University of Nottingham, University Park, Nottingham, NG7 2RD, UK\\
$^{13}$ Max-Planck-Institut für Astrophysik, Karl-Schwarzschild-Str. 1, D-85748 Garching b. München, Germany\\
$^{14}$ Centre for Astrophysics and Supercomputing, Swinburne University of Technology, P.O. Box 218, Hawthorn, VIC 3122, Australia\\
$^{15}$ Department of Physics, Indian Institute of Science, Bangalore 560012, India\\
}

\begin{abstract}

Neutral regions of the intergalactic medium (IGM) during the Epoch of Reionization (EoR) remain largely unexplored due to the limitations of existing probes. Owing to discoveries of numerous high-redshift radio-bright sources, the 21-cm forest, a series of absorption features imprinted by the neutral IGM in the spectra of such sources, now offers an attractive probe of the thermal and ionization state of the predominantly neutral IGM at $z\gtrsim5.5$. We analyse archival upgraded Giant Metrewave Radio Telescope (uGMRT) observations of J352-15, the brightest known radio-loud quasar in the EoR ($z=5.82$), to measure the one-dimensional (1D) power spectrum of the 21-cm forest. By comparing the observed power spectrum with forward-modelled synthetic spectra generated from cosmological simulations spanning a wide range of ionization and X-ray pre-heating scenarios, we perform Bayesian inference even in the absence of a statistical detection. We also present an independent Murchison Widefield Array measurement, although its lower sensitivity prevents competitive constraints. Using uGMRT, we achieve a sensitivity of $3.62\,\rm mJy\,beam^{-1}$ per $6.1\,\rm kHz$ channel. While we do not detect the 21-cm forest statistically, the null detection jointly constrains the mean neutral hydrogen fraction, $\langle x_{\rm HI}\rangle$, and the mean temperature of the neutral IGM, $\langle T_{\rm HI}\rangle$. At the $68\%$ credible level, our analysis disfavours cold and substantially neutral IGM models at $z\approx5.6$, including models with $\langle T_{\rm HI}\rangle \lesssim 27\,\rm K$ for $\langle x_{\rm HI}\rangle=0.1$. These limits probe parameter space allowed by existing Ly$\alpha$ and 21-cm observations, indicating substantial pre-heating of the neutral IGM above the adiabatic cooling floor. This demonstrates that the 21-cm forest has entered the regime of observationally informative statistics.

\end{abstract}

\vspace{0.3cm}

\noindent\footnotesize
$^{*}$ Corresponding author:
\texttt{tomas.soltinsky@inaf.it}

\vspace{0.5cm}
\thispagestyle{fancy}

\end{@twocolumnfalse}
]


\section{Introduction}

Over 12 billion years ago, the Universe underwent its last major phase transition in baryonic matter, evolving from a cold, predominantly neutral intergalactic medium (IGM) into a hot, highly ionised one. This era, called the Epoch of Reionization (EoR), dramatically reshaped the physical state of the IGM, setting the environmental conditions under which the first generations of stars, galaxies and black holes assembled and grew. The EoR is also relevant for precision cosmology: its imprint on the Cosmic Microwave Background (CMB) through Thomson scattering (and the associated optical depth and secondary anisotropies) is an unavoidable astrophysical contribution, and imperfect modelling can propagate into biases and degraded constraints on fundamental cosmological parameters. Constraining the timing and topology of reionization, together with the thermal and ionization history of the IGM, therefore remains one of the major open problems in modern cosmology.

Existing probes of the EoR, however, provide only limited information on the physical conditions within neutral regions of the IGM. The Ly$\alpha$ forest predominantly samples gas with only trace amounts of neutral hydrogen because Ly$\alpha$ absorption rapidly saturates in more neutral regions \citep[e.g.][]{Fan_2006,Yang_2020_z63}. As a result, it mainly constrains the late stages of reionization, once the mean transmission becomes non-negligible. Complementary information comes from Ly$\alpha$-emitter populations \citep[e.g.][]{Maitra_2025,Maitra_2026} and from Ly$\alpha$ damping-wing analyses of quasar spectra \citep[e.g.][]{Mason_2018,Kist_2025}, but these constraints are indirect, model-dependent, and often limited by small samples. Consequently, both the earlier phases of reionization and, crucially, the physical conditions inside the neutral regions of the IGM even at late times remain only weakly constrained. The situation is even more severe for the thermal state: traditional observables provide virtually no direct constraints on the temperature of the predominantly neutral IGM at $z \gtrsim 6$ \citep{Gaikwad_2023,Umeda_2025,Qin_2025}.

These gaps highlight the importance of complementary probes capable of directly accessing the neutral IGM. In particular, major experimental programmes are targeting the redshifted 21-cm signal from the EoR using the CMB as a radio background, including interferometric efforts such as Murchison Widefield Array \citep[MWA,][]{Bowman_2013_MWA}, The Low Frequency Array \citep[LOFAR,][]{vanHaarlem_2013_LOFAR}, Hydrogen Epoch of Reionization Array \citep[HERA,][]{DeBoer_2017_HERA}, the upgraded GMRT \citep[uGMRT,][]{Gupta_2017_uGMRT}, and ultimately the low frequency component of the Square Kilometre Array \citep[SKA-Low,][]{Dewdney_2009_SKA,SKA_Sciencbook_2026_CDEoR} as well as global-signal experiments such as the Experiment to Detect the Global Epoch of Reionization Signature \citep[EDGES,][]{Bowman_2018,Capallo_2025_EDGES}, the Mapper of the IGM Spin Temperature \citep[MIST,][]{Monsalve_2024_MIST}, Shaped Antenna measurement of the background Radio Spectrum \citep[SARAS,][]{Patra_2013_SARAS,Singh_2018_SARAS2,Nambissan_2021_SARAS3}, and the Radio Experiment for the Analysis of Cosmic Hydrogen \citep[REACH,][]{deLeraAcedo_2022_REACH}. While these initiatives have made impressive progress, the tomographic 21-cm emission experiments face a severe challenge from astrophysical foregrounds, mainly caused by synchrotron and free-free emission from the Milky Way as well as from extragalactic sources.  These foregrounds are brighter than the expected signal by orders of magnitude.  This in turn demands exquisite control of calibration, chromaticity, and spectral structure.  As a result, these experiments are so far limited to upper limits rather than detections; current observations disfavour the coldest IGM thermal histories, excluding spin temperatures below $T_{\rm S}\sim1.3-19\,\rm K$ at $z\geq6$ \citep{Greig_2021_LOFAR,Greig_2021_MWA,Hera_2023,Dhandha_2025}. 

These radio foreground limitations are alleviated by the 21-cm forest, which probes the EoR through narrow 21-cm absorption imprinted on the spectra of bright radio quasars \citep{Carilli_2002,Furlanetto_Loeb_2002,Ciardi_2013} or gamma-ray burst afterglows \citep{Ioka_2005,Ciardi_2015_GRB}, rather than through fluctuations against the CMB. By construction, the 21-cm forest is directly sensitive to neutral hydrogen along individual sightlines, in contrast to the Ly$\alpha$ forest, which becomes insensitive once the IGM is substantially neutral. Beyond this unique diagnostic power, a key practical advantage is that the forest is largely immune to the bright, spectrally smooth astrophysical foregrounds that dominate traditional 21-cm emission experiments: it is a differential absorption signal measured against a compact background source \citep{Furlanetto_Oh_2006,Pritchard_2012}. Because it probes the primordial IGM, the 21-cm forest has the potential to constrain the thermal state of the neutral IGM at high redshift \citep[e.g.][]{Xu_2009,Xu_2011,Mack_2012,Soltinsky_2021,Soltinsky_2025,Patil_2026}\footnote{Besides the 21-cm forest, only the \MgII \citep{Hennawi_2021,Tie_2024} and \OI \citep{Oh_2002,Keating_2014} lines have been studied as a probe of the neutral IGM, however not of its thermal state.} Furthermore, the 21-cm forest signal can probe the metal enrichment of the IGM \citep{Bhagwat_2022}, the growth of early supermassive black holes \citep{Soltinsky_2023}, the population of radio-loud active galactic nuclei \citep{Ewall_Wice_2014}, minihaloes \citep{Xu_2010,Meiksin_2011,Kadota_2023,Naruse_2024}, primordial black holes \citep{VillanuevaDomingo2022,Zhao_2026} and the microphysics of dark matter \citep{Shimabukuro_2014,Shimabukuro_2020,Shimabukuro_2023,Shimabukuro_2025_wavelets,Shimabukuro_2026_Topological,Shao_2023} and neutrinos \citep{Shao_2025_neutrino}. In addition, forest observables can help break degeneracies among cosmological and astrophysical parameters \citep[e.g.][]{Shao_2023,Shimabukuro_2025_wavelets,Shimabukuro_2026_Topological,Sun_2025}.

The 21-cm forest has not yet been detected, despite the attempts by \citet{Carilli_2007}, in which the main reason for non-detection was the low redshift of the targeted sources ($z=5.1$ and $5.2$). Nevertheless, several developments now make a renewed attempt timely and increasingly feasible. First, the sample of known radio-loud quasars (RLQSO) at $z>5.5$ -- viable background sources -- has grown rapidly since 2020, to $\sim 34$ targets \citep[e.g.][]{Belladitta_2020,Liu_2021,Banados_2021,Banados_2023,Banados_2025,Ighina_2021,Ighina_2023,Ighina_2024,Endsley_2023,Gloudemans_2022,Gloudemans_2023,Wolf_2024} \footnote{We maintain an up-to-date list of all known $z>5.5$ RLQSOs at \url{https://tomassoltinsky.github.io//eor/}}. In addition, significantly more high-$z$ RLQSO are expected to be detected \citep{Niu_2025}, especially when large radio surveys like LOFAR Two-metre Sky Survey \citep[LoTSS][]{Shimwell_2017,Kondapally_2021}, the Tata Institute of Fundamental Research (TIFR) Giant Metrewave Radio Telescope (GMRT) Sky Survey \citep[TGSS;][]{Intema_2017}, and the Galactic and Extragalactic All-sky MWA survey \citep[GLEAM;][]{Wayth_2015} are combined with observational programmes such as the William Herschel Telescope Enhanced Area Velocity Explorer (WEAVE)-LOFAR survey \citep{Smith_2016_WEAVE} and Euclid \citep{Euclid_2019}. Second, multiple lines of evidence favour a late end to reionization: the large sightline-to-sightline scatter in Ly$\alpha$ forest transmission disfavours a uniformly ionized IGM at $z\gtrsim 5.3$ \citep{Becker_2015,Kulkarni_2019,Bosman_2022}, the measurements of long dark gaps in the \Lya~forest \citep{Zhu_2021,Maity_2026} and \Lyb~forest \citep{Zhu_2022}, \Lya~forest transmission spikes at $z>5$ \citep{Gaikwad_2020,Nakane_2024}, clustering of \Lya~emitters \citep{Weinberger_2019}, deficit of \Lya~emitting galaxies around extended \Lya~absorption troughs \citep{Kashino_2020,Keating_2020,Christenson_2021}, mean free path of ionizing photons at $z=6$ \citep{Becker_2021,Cain_2021,Zhu_2023,Gaikwad_2023}, recent analyses of damping-wing absorption \citep{Becker_2024,Spina_2024,Zhu_2024,Sawyer_2025}, the observed upturn in the line density of \ion{O}{I} absorbers at $z\gtrsim5.5$ \citep{Becker_2019,Sebastian_2024}, and cross-correlation between [\ion{O}{III}] emitters and Ly$\alpha$ forest transmission at $z>5.4$ \citep{Kakiichi_2025} support the presence of neutral islands at $z<6$. Such late-reionization scenarios predict strong 21-cm forest absorption down to $z\simeq 6$ under favourable conditions \citep{Soltinsky_2021}. Third, new statistical observables including Wavelet Scattering Transforms \citep{Shimabukuro_2025_wavelets}, topological features \citep{Shimabukuro_2026_Topological} and -- notably the one-dimensional (1D) power spectrum of the forest \citep{Thyagarajan_2020, Shao_2023, Shao_2025_analytical, Soltinsky_2025} -- provide a route to detection without requiring line-by-line identification, substantially relaxing observational requirements. Furthermore, machine learning techniques can also reduce the observational requirements for the detection of the 21-cm forest signal \citep{Patil_2026}. For a more detailed review on the 21-cm forest see \citet{SKA_Sciencbook_2026_21cmForest}.


Within this context we analyse uGMRT archival observations of the brightest currently known RLQSO within the EoR, PSO J352.4034--15.3373 (hereafter J352--15). Since its discovery by \citet{Banados_2018}, J352--15 has been observed across radio, infrared, optical and X-ray wavelengths \citep[e.g.][]{Rojas-Ruiz_2021,Connor_2021}, with studies exploring its supermassive black hole accretion \citep{Rojas-Ruiz_2025} and radio jet orientation \citep{Walter_2025}. Early observational efforts to detect the 21-cm forest \citep[e.g.][]{Carilli_2007} focused primarily on searches for individual absorption features or broad optical-depth constraints. In contrast, our primary observable is the 1D power spectrum of the 21-cm forest absorption, a statistical observable designed to extract information even when individual absorption features are not directly detectable. Building on the framework developed in \citet{Soltinsky_2025}, we show that even a null detection of the 21-cm forest signal can place meaningful constraints on the thermal and ionization state of the predominantly neutral IGM during the final stages of reionization. While our analysis results in no statistical detection of the signal, we demonstrate that current-generation 21-cm forest observations have entered the regime of observationally informative constraints, particularly disfavouring very cold and substantially neutral IGM models. This is in agreement with 21-cm tomography observations at $z>6.5$ by LOFAR \citep{Greig_2021_LOFAR,Ghara_2025}, MWA \citep{Greig_2021_MWA} and HERA \citep{Hera_2023}.

This study is structured as follows. We start by describing the uGMRT archival data and its reduction into the radio spectrum of J352--15 in Sec.~\ref{sec:observations}. In Sec.~\ref{sec:mocks} we then describe how we model the mock 21-cm forest 1D power spectrum observations which is then used in IGM properties inference described in Sec.~\ref{sec:constraints}. Forecasted observations and parameter inference based on them are presented in Sec.~\ref{sec:constraints_forecasts}. We summarize and conclude our findings in Sec.~\ref{sec:conclusions}.

\section{uGMRT observations of J352--15}\label{sec:observations}

In this section we motivate our selection of J352--15 as the target RLQSO, describe the archival uGMRT observations and their reduction, and present the resulting radio spectrum and 1D power spectrum.

\subsection{Target selection}\label{sec:target_selection}

\begin{figure*}
\vspace{-0.5cm}
   \centering 
   \includegraphics[width=1\linewidth, keepaspectratio, trim = 0 0 0 0]{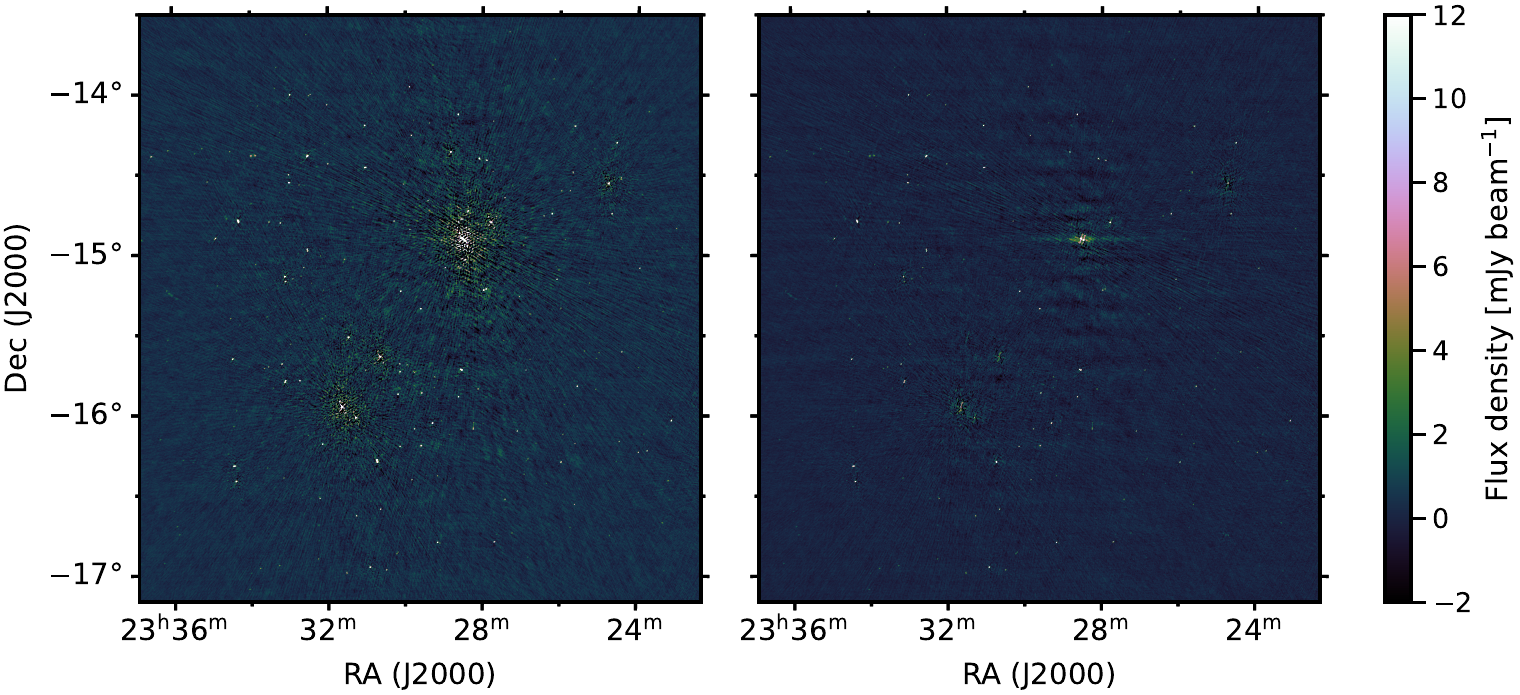}
\caption{The continuum image of the target field at $212.75\,\rm MHz$, showing the central $3.5\degree \times 3.5\degree$ area. The left panel shows the image after direction-independent self-calibration. The off-source rms noise near the field center, away from the bright source, was about $1.5\,\rm mJy\,beam^{-1}$ with a synthesized beam width of about $11''$. The right panel shows the image after direction-dependent calibration, which reaches an rms of about $\sim0.4\,\rm mJy\,beam^{-1}$.  We observe an improvement in image quality after direction-dependent calibration, with a markedly reduced direction-dependent artifact around bright off-axis sources. However, residual artifacts remain around the brightest off-axis source in the field (top right), indicating that direction-dependent calibration errors  still contribute to the image noise.}
\label{fig:cont_map}
\end{figure*}

J352--15 is a radio-loud quasar located at ${\rm RA}=23^{\rm h}\,29^{\rm m}\,36.825^{\rm s}$, ${\rm Dec}=-15^\circ\,20^\prime\,14.414^{\prime\prime}$, with redshift $z=5.82$, discovered by \citet{Banados_2018}. The field of view (FoV) centred at this RLQSO is shown in Fig.~\ref{fig:cont_map}. It was selected as a high-$z$ candidate by criteria based on drop-in fluxes and its flat optical continuum based on observations within Panoramic Survey Telescope \& Rapid Response System 1 (Pan-STARRS1). This was later confirmed by a follow-up spectroscopic observation by the Low-Dispersion Survey Spectrograph on the Magellan Clay telescope. While the presence of radio source in the vicinity of J352-15 was already measured in the TGSS, GLEAM and 1.4 GHz NRAO VLA Sky Survey \citep[NVSS;][]{Condon_1998}, the fact that this radio emission was coming from this quasar was confirmed by the observations with the Karl G. Jansky Very Large Array (VLA) \citep{Banados_2018}.

Its flux density at $150\,\rm MHz$ is $S(\nu=150\,\rm MHz)=110.6\pm13.8\,\mathrm{mJy}$ as measured in TGSS. Assuming that the radio spectrum continuum is a single power-law with the measured radio spectral index $\alpha_{\rm R}=-0.89$ \citep{Banados_2018,Rojas-Ruiz_2021} the flux density at the target (redshifted) 21-cm frequency, $\nu_{\rm T}\simeq 208\,\mathrm{MHz}$, is $S(\nu=\nu_{\rm T})=81.7\,\mathrm{mJy}$, making it the brightest known RLQSO at $z>5.5$ (i.e. before reionization has completed). Therefore, it is the highest-signal-to-noise (SNR) background source currently available for 21-cm forest studies. Absorption is expected only bluewards of $\nu_{\rm T}$ (i.e.\ at $\nu>\nu_{\rm T}$), whereas the redward side of the spectrum ($\nu<\nu_{\rm T}$) should contain no forest signal. 

Hence, if the IGM is largely ionized and heated by $z\sim 5.4$ \citep{Kulkarni_2019,Keating_2020,Bosman_2022,Davies_2026}, corresponding to $\nu\simeq 222\,\mathrm{MHz}$, then one expects 21-cm forest absorption over a usable bandwidth of $\Delta\nu \approx 14\,\mathrm{MHz}$. This broad frequency interval samples a long line-of-sight path through the IGM, providing access to a large range of spectral modes for our statistical analysis. This corresponds to $\sim 200\,\mathrm{cMpc}$, significantly larger than the expected size of the quasar proximity region, which may suppress the 21-cm forest signal\footnote{Assuming that the quasar was accreting for $10\,\mathrm{Myr}$ and has a magnitude at $1450\,\mathrm{\mathring{A}}$ of $-25.59\pm0.13$ \citep{Banados_2018}, the quasar proximity zone can extend to $\sim66\,\mathrm{cMpc}$ \citep{Soltinsky_2023}.}. In addition, absorption associated with the host galaxy may also be present. The search for it is a ongoing work and will be presented in a future publication. Given that we are interested in the 21-cm forest absorption arising from the IGM and we do not model these proximate effects, we exclude the $208\,\mathrm{MHz}-209\,\mathrm{MHz}$ region from our analysis.

Although higher-redshift RLQSOs, such as recently discovered J0410--0139 at $z\sim7$ \citep{Banados_2025}, would in principle offer a longer spectral lever arm, they are significantly fainter and hence yield lower sensitivity for the forest measurement. In addition, optical spectroscopy of J352--15 by the Low-Dispersion Survey Spectrograph on the Magellan Clay telescope reveals a pronounced Gunn–Peterson trough blueward of Ly$\alpha$ together with evidence for a damping-wing absorption profile, indicating substantial neutral hydrogen along the line of sight (LOS) \citep{Banados_2018}. Although these observations do not directly imply detectable 21-cm absorption, they suggest that J352--15 probes an unusually neutral environment compared with many other known RLQSOs at similar redshifts. This further motivates J352--15 as one of the most promising current targets for 21-cm forest observations.  

\subsection{Usable spectral modes and sidelobe avoidance}\label{sec:sidelobes}

Before discussing the accessible spectral modes, we first clarify the notation used for Fourier variables. Throughout this work, we use $\kappa_\nu$ to denote the angular Fourier conjugate of frequency, with units of $\rm MHz^{-1}$. This quantity is related to the Fourier-delay variable, $\eta$, commonly used in 21-cm analyses by $\kappa_\nu = 2\pi\eta$. We reserve $k_\parallel$ for the cosmological LOS wavenumber, expressed in $h\,{\rm cMpc}^{-1}$. The two are related by $k_\parallel = \kappa_{\nu}/Y(z)$, where $Y(z)=\mathrm{d}r_\parallel/\mathrm{d}\nu$ converts frequency intervals into comoving LOS distances. Here $r_\parallel$ denotes the comoving radial distance. 

We estimate the spectral modes expected to be contaminated by sidelobe-induced mode mixing following Eq.~30 of \citet{Thyagarajan_2020}. This equation defines the boundary of the foreground ``wedge'' in $(k_\perp, k_\parallel)$ space, which arises from chromatic leakage of foreground emission due to imperfect subtraction of off-axis sources and the chromatic instrumental response. The extent of the wedge is set by the maximum geometric delay associated with emission within the primary beam, and is therefore primarily determined by the effective FoV and baseline length.

Since our scientific analysis is based on the 1D (LOS) power spectrum of the normalized quasar spectrum, we conservatively define a characteristic spectral wavenumber $k_{\rm SL}$ corresponding to the wedge boundary evaluated at the largest transverse mode probed by the data, i.e. at $k_{\perp,\max}$ set by the synthesized beam and the adopted $uv$ weighting. Modes with $k_\parallel < k_{\rm SL}$ are conservatively treated as potentially affected by sidelobe leakage and residual foreground structure, and are excluded from the analysis. We retain only modes with $k_\parallel \geq k_{\rm SL}$, which lie in the ``EoR window'' where contamination from mode-mixing is expected to be low.

Assuming a $\Lambda$CDM cosmology, the uGMRT full width at half maximum (FWHM) of the synthesized point-spread function (PSF) $\theta_{\rm S}=15.4''$ and FWHM of the primary beam of the antenna power pattern $\theta_{\rm P}=1.8\degree$\footnote{Based on The GMRT: System Parameters and Current Status from cycle 50 - \href{http://indrayani.ncra.tifr.res.in/~secr-ops/sch/c50webfiles/gtac_50_status_doc.pdf}{http://indrayani.ncra.tifr.res.in/$\sim$secr-ops/sch/c50webfiles/gtac\_50\_status\_doc.pdf} considering $\nu=208\,\rm MHz$.} results in $k_{\rm SL}=0.634\,h\rm cMpc^{-1}$ which corresponds to $\kappa_{\nu,\rm SL}=1\,\rm MHz^{-1}$. We therefore exclude modes with $\kappa_{\nu} < \kappa_{\nu,\rm SL}$ from the analysis. 

Note that while MWA has a shorter maximum baseline ($B_{\rm max}=3\,\rm km$), and hence a larger $\theta_{\rm S}$, it also has a much larger FoV ($\theta_{\rm P}=18.8\degree$). On the other hand, LOFAR EoR observations have a $\theta_{\rm P}$ comparable to that of the uGMRT but typically employ the Dutch core and remote stations, corresponding to a maximum baseline of $B_{\rm max}=100\,\rm km$ \citep{vanHaarlem_2013_LOFAR}. Although the full LOFAR array includes international stations extending to baselines exceeding $1000\,\rm km$, these are not typically used for EoR analyses. Using these nominal instrumental parameters, the corresponding $\kappa_{\nu,\rm SL}=1.6\,\rm MHz^{-1}$ and $7.4\,\rm MHz^{-1}$ for MWA and LOFAR, respectively. This comparison illustrates that the combination of angular resolution and FoV of the uGMRT results in a relatively low characteristic sidelobe-contamination scale. This makes the uGMRT particularly well suited for 21-cm forest power-spectrum studies, since lower-$k_\parallel$ modes are expected to provide higher SNR \citep{Soltinsky_2025}. We note, however, that the effective range of usable modes in a given analysis also depends on the adopted baseline selection and foreground-modelling strategy.

The highest $\kappa_{\nu}$ accessible is dictated by the frequency resolution. For example, the frequency channel width of $\delta\nu=6.1\,\rm kHz$ corresponds to a maximum accessible spectral mode of $\kappa_{\nu}\approx515\,\rm MHz^{-1}$. However, even in the most optimistic models the signal is dominated by thermal noise at high $k_\parallel$ \citep{Soltinsky_2025}. 

\subsection{Observational data calibration}\label{sec:calibration}

In what follows we have analysed archival uGMRT data combining ddtC007 (PI: Chris Carilli) and ddtC219 (PI: Arnab Chakraborty) proposals, both targeting J352--15, with a total on-source time of $t_{\rm int}=17.5\,\rm hr$ observed by Band-2 receivers. This data was obtained over two nights in 2018 and three nights in 2022 as shown in Table~\ref{table:t_int}. Even though the goals of these proposals were to explore the systematics, Radio Frequency Interference (RFI) environment, study the J352-15's spectral properties and search for \HI~absorption associated with the host galaxy, we use this archival data in the context of 21-cm forest arising from the diffuse IGM along the LOS. 

\begin{table*}[ht!]
\caption{Total and effective integration time during different observing runs and corresponding mean noise RMS after adding the data from the particular run.}                 
\label{table:t_int}    
\centering                        
\begin{tabular}{c c c c c c}      
\hline\hline               
Date & Proposal ID & $t_{\mathrm{int}}\,\mathrm{[hr]}$ & $t_{\mathrm{eff}}\,\mathrm{[hr]}$ & Fraction excised & $\sigma_{\rm N}\,[\rm mJy\,beam^{-1}\,per\,channel]$ \\         
\hline                      
 24.6.2018 & ddtC007 & 2.77 & 1.66 & $40.08\%$ & 9.63 \\
 30.6.2018 & ddtC007 & 3.10 & 1.99 & $35.68\%$ & 5.63 \\
 17.7.2022 & ddtC219 & 6.00 & 4.36 & $27.32\%$ & 4.71 \\
 27.8.2022 & ddtC219 & 2.75 & 2.34 & $14.82\%$ & 4.13 \\
 28.11.2022 & ddtC219 & 2.88 & 1.37 & $52.44\%$ & 3.62 \\
\hline  
 Total & Both & 17.50 & 11.73 & $32.99\%$ & 3.62 \\
\hline  
\end{tabular}
\end{table*}

A bandwidth of $25\,\rm MHz$, subdivided into 16384 channels with a resolution of $1.5\,\rm kHz$ ($2\,\rm km/s$), was used for the observation with GMRT wideband backend (GWB) as the correlator. The integration time per visibility point was $8\,\rm s$. The frequency bandwidth of the data observed in 2018 (ddtC007) covers 203–228 MHz, whereas that observed in 2022 (ddtC219) covers 197.5–222.5 MHz. We analysed each dataset separately using the same procedure described below, and then produced a combined image from all nights over the common band shared by the two, 203–222.5 MHz. The actual bandwidth of the final image is therefore about 19.5 MHz. The standard calibrator 3C~48 was observed to calibrate the flux density scale, while regular observations of the nearby compact source 2321-163 were used to calibrate the complex antenna gains.


According to the Exposure Time Calculator (ETC) for the upgraded Giant Metrewave Radio Telescope handbook\footnote{\href{http://www.ncra.tifr.res.in:8081/~secr-ops/etc/etc_help.pdf}{http://www.ncra.tifr.res.in:8081/$\sim$secr-ops/etc/etc\_help.pdf}}, the frequencies rendered unusable by RFI lie below our frequencies of interest, being most severe in the 165–190 MHz range. Even so, RFI remains a concern within our band and can corrupt the data significantly. The data were first inspected using the \textsc{AOFLAGGER}\footnote{\href{https://aoflagger.readthedocs.io/en/latest/}{https://aoflagger.readthedocs.io/en/latest/}} package for the detection and excision of RFI \citep{Offringa_2012}. We then employed a {\sc casa}-based flagging and calibration pipeline to solve for the complex gains and remove any remaining bad data, following standard procedure.
The autoflagging algorithms {\sc tfcrop} and {\sc rflag} in {\sc casa}'s {\sc flagdata} task were used to identify and excise further RFI. The flux density of the primary calibrator, 3C~48, was set using the Perley-Butler model \citep{Perley_2017}. The primary calibrator was used to derive delay and bandpass
solutions, and frequency-independent complex gain corrections using the {\sc gaincal} and {\sc bandpass} tasks in {\sc casa}.  The time-variable complex gains for each antenna were obtained from the secondary calibrator 2321--163, observed for 6min in every 30-min scan of the target. The resulting calibration solutions were applied to the target field, which was then split off for imaging and self-calibration. At this stage we averaged the data by 2 channels, yielding a $\sim 3.05\,\rm kHz$  spectral resolution, while retaining the full time resolution. This helped us identify and flag bad data during the self-calibration and imaging loops.


We used {\sc wsclean} \citep{Offringa_2014} to make the continuum image of the target field. We used multi-scale wide-band deconvolution together with auto-masking \citep{Offringa_2017} to capture the variation of sky brightness across this large bandwidth and over different spatial scales. The full band was first imaged without a cleaning mask, with deconvolution terminated after 50k iterations. From the resulting image, we derived a clean mask by excluding the region below a local threshold of 6$\sigma$, where $\sigma$ is an estimate of local RMS, following the method of \citet{Tasse2021}. We then ran a second, constrained deconvolution with this mask to generate an artifact-free model for self-calibration. A region of $\sim 4.5\degree \times 4.5\degree$ was imaged with $8192 \times 8192$ pixels of $2''$ each, extending into the sidelobes of the primary beam at these frequencies so that bright off-axis sources could be properly deconvolved.

{\sc wsclean} inverted the frequency-dependent sky model derived from the deconvolution process into model visibilities at the end of the imaging process, which we used for self-calibration. We performed several rounds of phase-only self-calibration, with an improved mask at each iteration, until no further improvements were seen in the continuum image. We used Briggs weighting with a robust parameter of $-1$ during self-calibration \citep{Briggs_1995}. The continuum image after direction-independent calibration is shown in the left panel of Fig.\ref{fig:cont_map}. The off-source RMS noise near the field center, away from the bright source, was about $1.5\,\rm mJy\,beam^{-1}$ with a synthesized beam width of about $11''$.

We found that there were direction-dependent artifacts around 3 bright sources away from the phase center. Traditional direction-independent calibration was unable to remove the error patterns around these off-axis sources. We followed the same procedure as described in \citet{Heywood_2020, Heywood_2022} to solve for direction-dependent effects. First, we used \textsc{DDFacet}\footnote{\href{https://github.com/saopicc/DDFacet}{{https://github.com/saopicc/DDFacet}}} \citep{Tasse_2018} to make an image of the target field. In \textsc{DDFacet}, the entire field was divided into facets (here 8) along which deconvolution took place, and the direction-dependent model for each facet was stored. An additional all-sky model data column was created using the predict option of \textsc{wsclean} for the entire sky. The differential gain method \citep{Oleg_2011} was then employed for direction-dependent calibration using \textsc{CUBICAL} \citep{Kenyon_2018}. Traditional complex gains (G) were computed based on an all-inclusive sky model. Additionally, complex differential gain terms (dE) were derived for problematic sources using the model predicted on the fly from the \textsc{DDFacet} model output. The resultant image after direction-dependent calibration is shown in the right panel of Fig. \ref{fig:cont_map}. The off-source rms, after direction-dependent calibration, was $\sim0.4\,\rm mJy\,beam^{-1}$. Although the improvements after the direction-dependent calibration are clearly visible in Fig. \ref{fig:cont_map}, we still observed some residual direction-dependent artifacts around the brightest off-axis source and low-level residual RFI (top right in Fig. \ref{fig:cont_map}).  We are actively working on improving the RFI flagging and on developing a more robust direction-dependent calibration pipeline, which we defer to future work.

After direction-dependent calibration, the continuum emission was subtracted from the calibrated multi-channel visibilities using the {\sc uvsub} routine in {\sc casa}. Any residual continuum emission was subtracted via a 2nd-order polynomial fit to each visibility spectrum, and the residual visibilities were shifted to the barycentric frame using the {\sc mstransform} routine in {\sc casa}. We also averaged the data by 2 channels to reduce the data volume and improve the SNR, resulting in a spectral resolution of $\sim 6.1\,\rm kHz$. We then used {\sc casa} task {\sc tclean} to make a spectral image cube from the continuum-subtracted visibilities. The cube was produced in the barycentric frame with w-projection and natural weighting. We opted for natural weighting as it gives the lowest thermal noise, at the expense of angular resolution, which is not critical for our line-of-sight analysis. This yields a per-channel RMS of about $5\,\rm mJy\,beam^{-1}$. The synthesized beam size of the cube is around $\sim 18''$. We note that at this stage the spectra still contain residual broadband bandpass structure. The modelling and subtraction of these features, and the corresponding reduction in noise, are described in Sec.~\ref{sec:bandpass_subtraction}.

\begin{figure}
   \centering \includegraphics[width=1\linewidth, keepaspectratio, trim = 0 0 0 0]{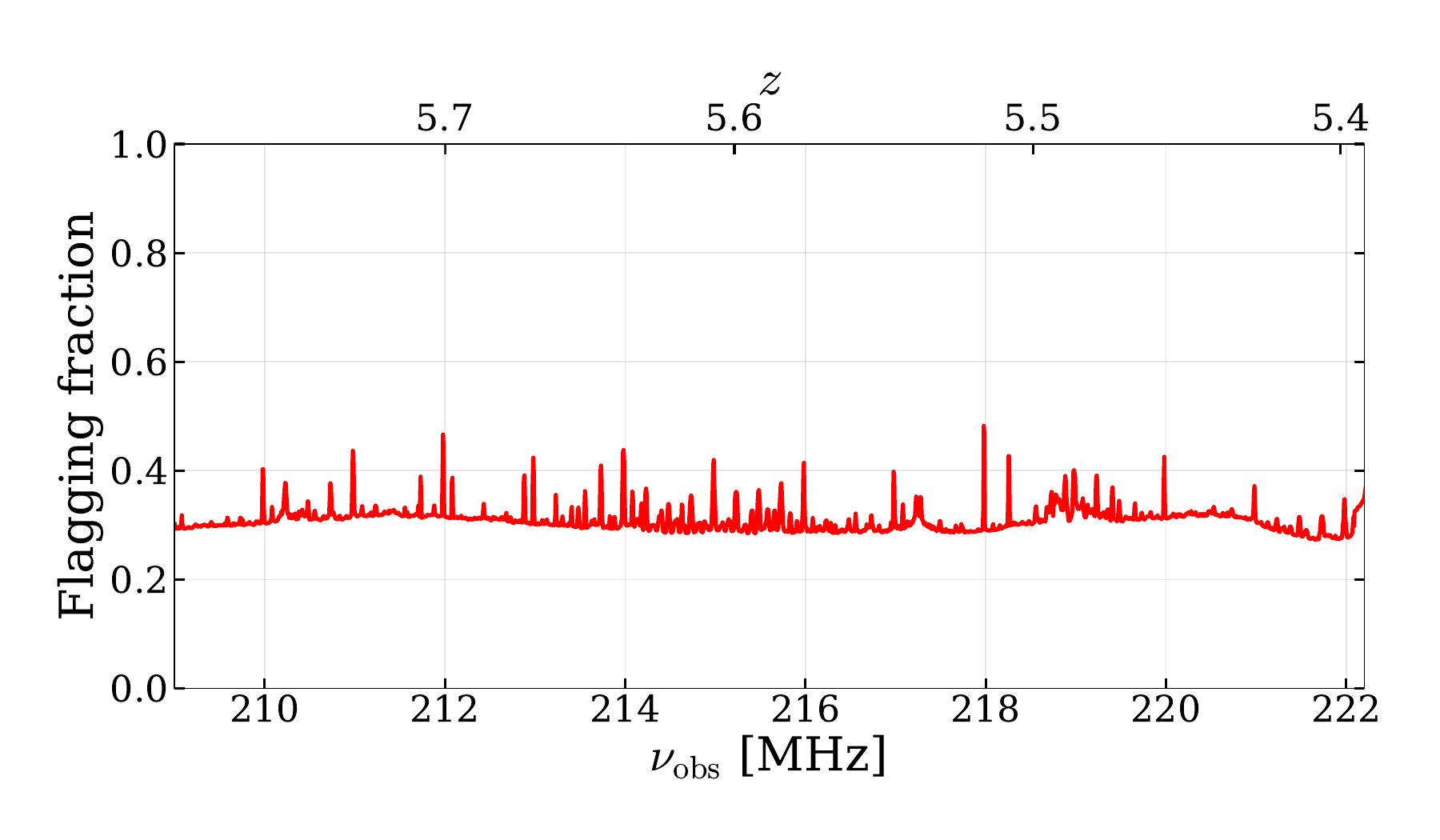}
\vspace{-0.6cm}
\caption{The fraction of data excised as a function of frequency, averaged over all observing nights, for the entire 17.5 hours of on-source time.}
\label{fig:flag_frac}
\end{figure}

The RFI environment and the direction-dependent ionospheric fluctuations varied between different nights of observations. The fraction of target-source data lost due to time-variable issues was $\sim 20$--$60\%$ between different nights as presented in App.~\ref{app:RFI_bynight}. The fraction of data excised as a function of frequency, averaged over all observing nights, for the entire 17.5 hours of on-source time is shown in Fig.~\ref{fig:flag_frac}. The overall fractional data lost across the band for each night and in total is summarized in Table~\ref{table:t_int}. For all nights combined it was around $33\%$ leaving us with the effective integration time of $t_{\rm eff}=11.73\,\rm hr$.

\subsection{Residual broadband features subtraction}
\label{sec:bandpass_subtraction}
Finally, the spectra exhibit broadband wave-like structure, most likely arising from residual bandpass features. An example of such features, particularly in the radio spectrum of J352--15, is shown in Appendix~\ref{app:flagging}. To address this issue, previous \HI~studies modelled the bandpass using low-order polynomials or Savitzky-Golay smoothing filters \citep{Gupta_2021,Heywood_2024}. However, these approaches require substantial fine-tuning, for example in the choice of polynomial order or smoothing scale. 

Instead, we model the residual bandpass using a discrete wavelet decomposition. Wavelet decomposition has previously been applied to astronomical spectra to isolate structures on different characteristic scales and suppress unwanted large-scale or noise-like components \citep[e.g.][]{Fligge_1997,Machado_2013,Jiang_2017}. Specifically, we employ a Daubechies-8 wavelet, which provides a flexible and localized representation of smooth spectral variations. We reconstruct only the approximation (coarsest-scale) component of the wavelet decomposition and treat it as the residual bandpass model. Subtracting this component removes broadband spectral structure while largely preserving small-scale fluctuations. The choice of retaining only the coarsest-scale component was made conservatively to minimise the risk of suppressing potential 21-cm forest features. This approach avoids the need to explicitly choose a polynomial order or smoothing scale and therefore reduces the degree of manual fine-tuning required.


\subsection{Empirical noise level estimation}

The radiometer equation predicts the noise RMS per channel of \citep[cf.][]{Datta_2007,Ciardi_2013}
\begin{equation}
    \sigma_{\rm N} = \left(\frac{A_{\rm eff}}{T_{\rm sys}}\right)_{N_{\rm d}=1}^{-1}\frac{\sqrt{2}k_B}{\sqrt{N_{\rm d}(N_{\rm d}-1)\Delta\nu t_{\rm eff}}}, \label{eq:radiometer}
\end{equation}
\noindent
where $\left(A_{\rm eff}/T_{\rm sys}\right)_{N_{\rm d}=1}$ is the frequency-dependent sensitivity for a single dish and $N_{\rm d}$ is the number of dishes. However, this does not encompass all realistic systematics many of which (e.g. RFI, ionospheric effects, and weather conditions) are difficult to predict. We therefore empirically infer the scaling of $\sigma_{\rm N}$ with $t_{\rm eff}$ directly from the archival data.

\begin{figure}
   \centering \includegraphics[width=1\linewidth, keepaspectratio, trim = 0 0 0 0]{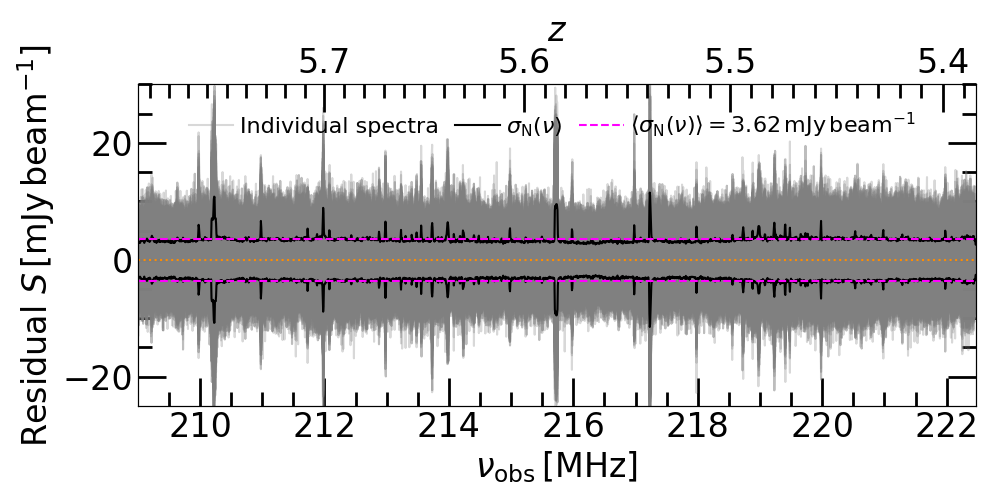}
\vspace{-0.6cm}
\caption{The spectra along 1000 arbitrary LOS of the continuum-subtracted image cube (in grey). The $1\sigma$ uncertainty of these spectra per channel is shown by the solid black curves. The average spectral RMS is approximately $3.62\,\rm mJy\,beam^{-1}$ per frequency channel and is shown by the dashed pink line.} 
\label{fig:rms}
\end{figure}

After completing all data calibration steps described in the previous sections, including the residual bandpass subtraction, we estimate $\sigma_{\rm N}$ by selecting 1000 randomly selected sightlines within the FoV close to the target J352--15. In Fig.~\ref{fig:rms} we show the observed residual flux density, $S$, spectra along 1000 arbitrary directions of the continuum-subtracted image cube, and the $1\sigma$ uncertainty of these spectra is shown by the solid black curve. The mean spectral RMS is $\sigma_{\mathrm{N}}=\langle\sigma_{\mathrm{N}}(\nu)\rangle=3.62\,\rm mJy\,beam^{-1}$ per frequency channel and is shown by the dashed pink line. 

We also took an off-source region of size $8$ times the synthesized beam close to the phase center and estimated the RMS for this region along the frequency axis. The resulting RMS is consistent with the estimate obtained from the randomly selected sightlines. Given the fraction of data lost due to RFI ($\sim 33\%$), the expected theoretical spectral RMS noise is around $2\,\rm mJy\,beam^{-1}$ per frequency channel according to the ETC. The measured spectral RMS noise is $\sim80\%$ higher than the theoretical expectation. We attribute this discrepancy primarily to residual artifacts around bright off-axis sources that contaminated the central part of the image through sidelobes, which resulted in higher RMS. 

\begin{figure}
    \begin{minipage}{\linewidth}
 	  \centering
 	  \includegraphics[width=\linewidth]{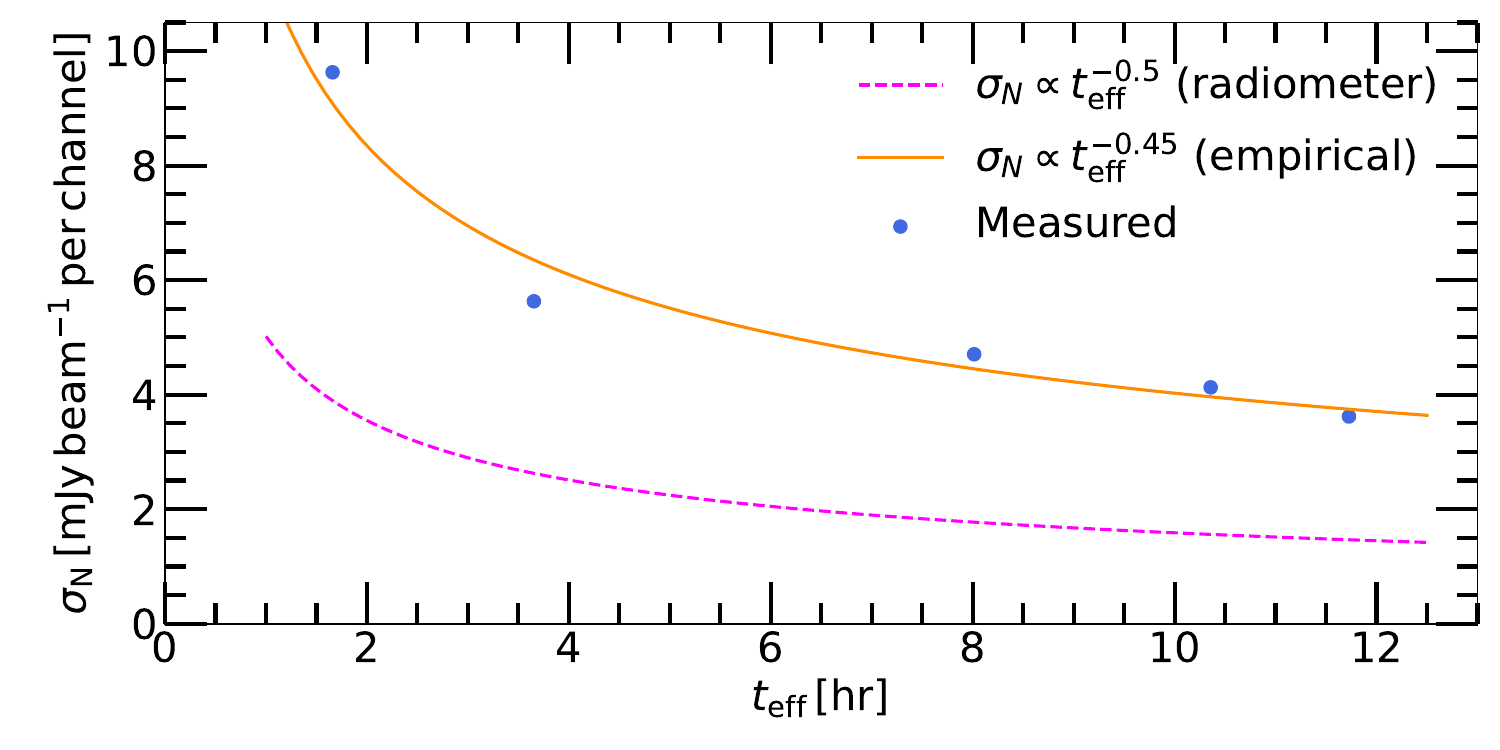}
	\end{minipage}
    \vspace{-0.3cm}
    \caption{The noise RMS measured for various combined nights from ddtC007 and ddtC219 (blue dots). A best-fit scaling of the noise RMS with the effective integration time (i.e. after RFI excision) is shown as solid orange curve. For comparison, we show the best-fit radiometer scaling (dashed pink curve).}
    \label{fig:rms_plot}
\end{figure}

We repeat this analysis for four different effective integration times. We start with data from a single night, particularly on 28.11.2022, and gradually add data from other nights. The resulting $\sigma_{\rm N}$ are shown as blue datapoints in Fig.~\ref{fig:rms_plot}. These datapoints are well fitted by a scaling relation
\begin{equation}
    \sigma_{\mathrm{N}}\propto t_{\mathrm{int}}^{-0.45}\label{eq:sigma_N}
\end{equation}
shown as solid orange curve. For comparison we also show the radiometer scaling from Eq.~\ref{eq:radiometer} (i.e. $\sigma_{\rm N}\propto t_{\rm eff}^{-0.5}$, dashed pink curve). Note that the amplitude is based on the uGMRT's ETC output. In practice, the noise decreases more slowly with added observational time than the theoretical prediction and is $\sim2-3$ times higher in amplitude.

\subsection{J352--15 radio spectrum}\label{sec:obs_spectrum}

\begin{figure*}
    \begin{minipage}{\linewidth}
 	  \centering
 	  \includegraphics[width=\linewidth]{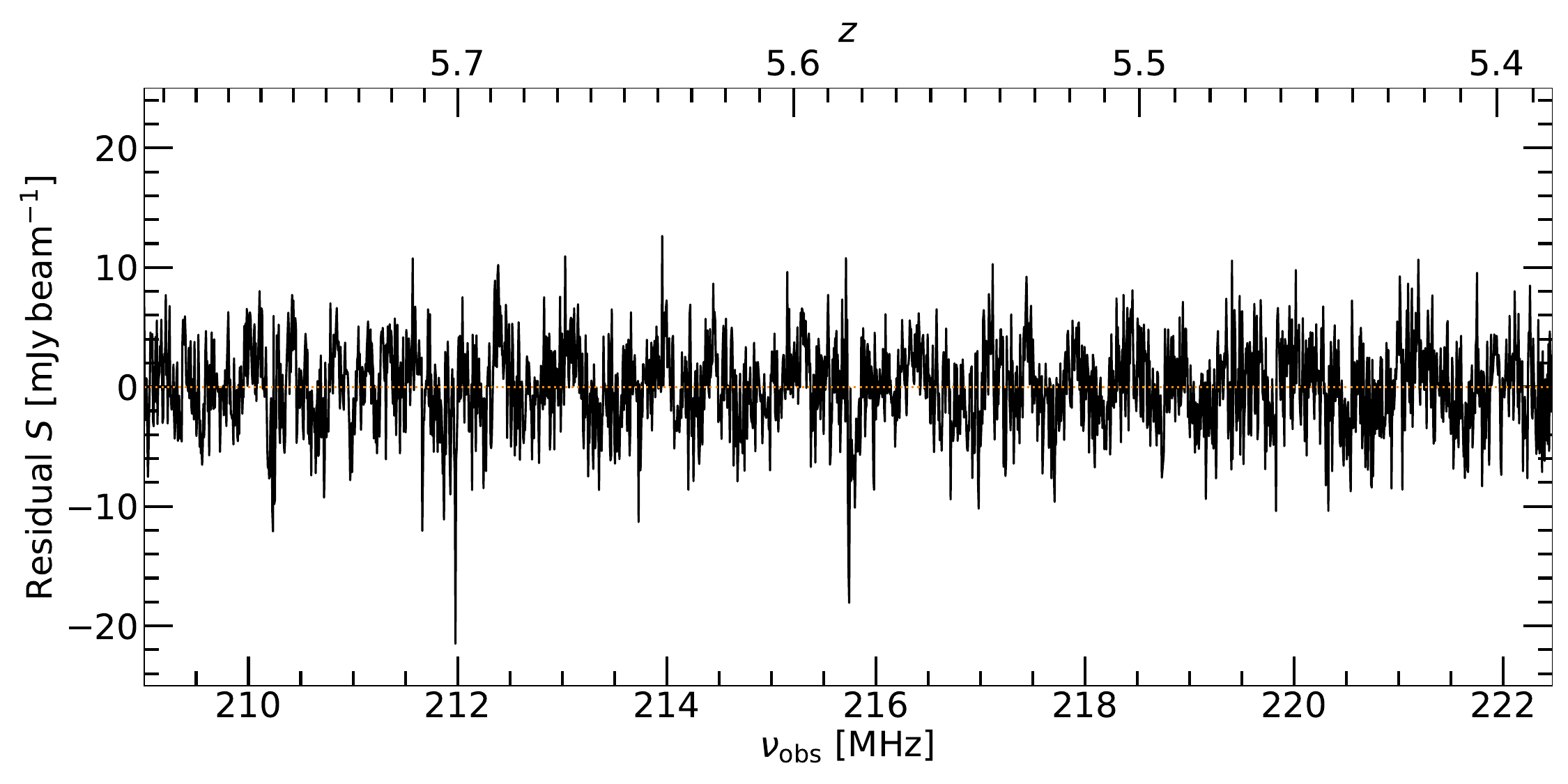}   
	\end{minipage}
    \vspace{-0.3cm}
    \caption{Residual flux-density spectrum of J352--15 after the calibration and continuum subtraction. Note that the strong absorption features (e.g. at $\sim210.3\,\mathrm{MHz}$, $\sim212.0\,\mathrm{MHz}$, $\sim213.7\,\mathrm{MHz}$ and $\sim215.7\,\mathrm{MHz}$) are in flagged channels which exhibit noise rms that is 1.5 times above mean $\sigma_{\rm N}$.} 
    \label{fig:F21_obs}
\end{figure*}

Here we present our measurement of the residual flux density spectrum of J352--15 at $209.0\,\mathrm{MHz}<\nu<222.5\,\mathrm{MHz}$ ($5.4\leq z\leq5.82$) in Fig.~\ref{fig:F21_obs}. The residual spectrum is dominated by instrumental noise. There are strong absorption features at $\sim210.3\,\mathrm{MHz}$, $\sim212.0\,\mathrm{MHz}$, $\sim213.7\,\mathrm{MHz}$ and $\sim215.7\,\mathrm{MHz}$, however, these are located at frequency channels that are strongly affected by systematic effects across many off-source spectra as shown in Fig.~\ref{fig:rms}. We therefore do not interpret these features as astrophysical in origin. Instead we identify peaks in $\sigma_{\mathrm{N}}(\nu)$ which have values $>1.5\sigma_{\mathrm{N}}$ (solid black curve vs dashed pink line in Fig.~\ref{fig:rms}), flag the surrounding 20 pixels (i.e. $\sim120\,\mathrm{kHz}$) and do not use them for the following analysis. These spectral channels are shown in App.~\ref{app:flagging}. We note that an astrophysical 21-cm absorption feature coincident with a masked channel would also be removed by this procedure. However, because the same channels are excluded from both the observations and forward-modelled spectra, the following analysis remains self-consistent and conservative.

\begin{figure}
    \begin{minipage}{1.\linewidth}
 	  \centering
 	  \includegraphics[width=\linewidth]{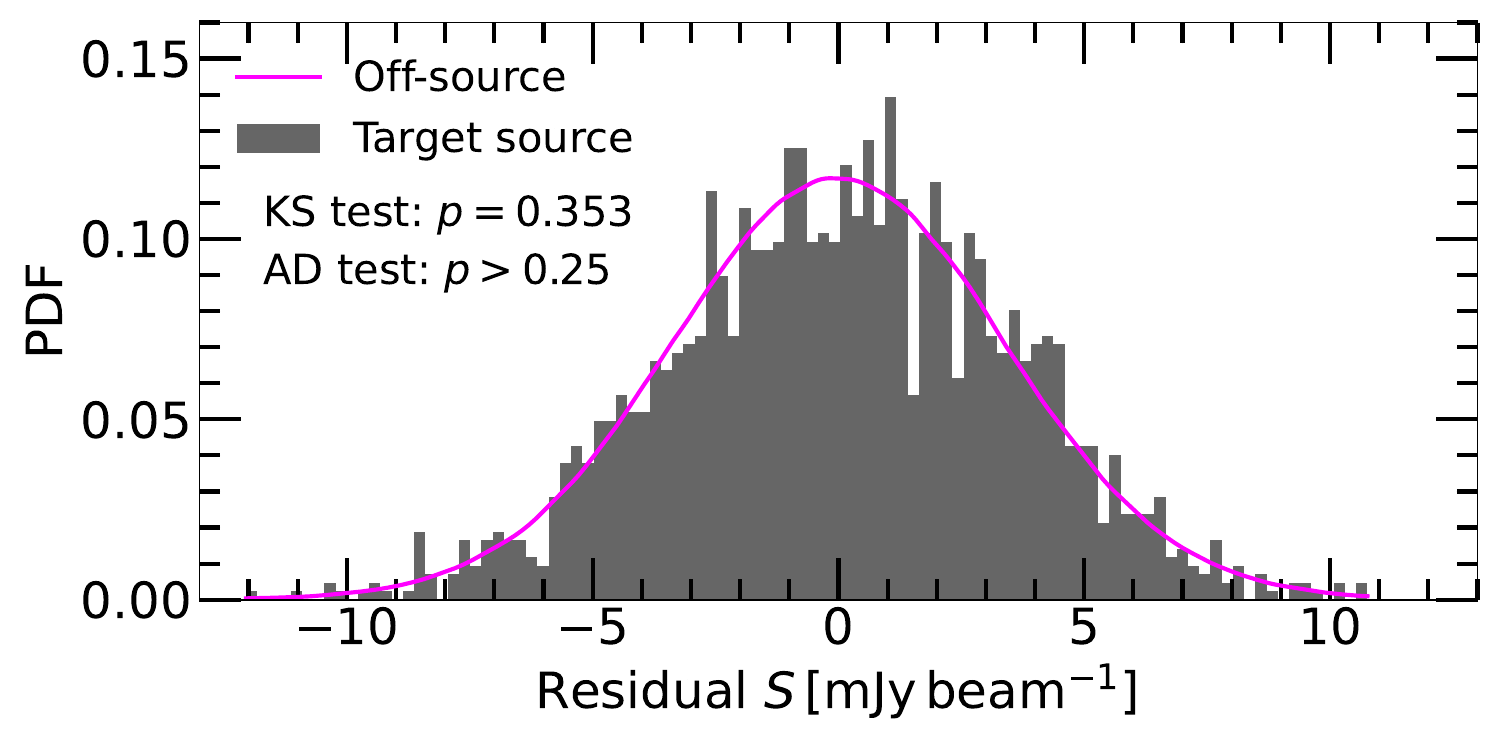}
	\end{minipage}
	\vspace{-0.3cm}
    \caption{Distribution of frequency-channel flux densities of the measured spectrum of J352--15 (grey bars) and off-source spectra (pink curve). According to both the KS and AD test, there is no significant evidence that these are drawn from different distributions.}
    \label{fig:flux_hist}
\end{figure}

We perform a similar test to the one presented in \citet{Soltinsky_2025}, where the statistical detection of the 21-cm forest can be confirmed by comparing the absorption (i.e. negative) part of the channel-by-channel flux density distribution of the noisy signal with noise-only spectra. However, instead of assuming Gaussian noise-only spectra, we use the empirically measured off-source spectra as the noise reference. Fig.~\ref{fig:flux_hist} shows this test, namely the target spectrum distribution (grey bars) overplotted by the distribution from the off-source spectra (solid pink curve). Performing both a Kolmogorov–Smirnov (KS) and Anderson-Darling (AD) tests we do not find a statistically significant difference between the distributions (KS $p$-value is 0.353, while for AD it is $>0.25$). We find no statistically significant excess on the absorption (negative-residual flux) side of the target-spectrum distribution. Instead, the observed distribution is broadly consistent with Gaussian white noise centred around zero continuum residual, with an RMS of $3.65\,\mathrm{mJy\,beam^{-1}}$ (i.e. close to the measured $\sigma_{\mathrm{N}}$). We therefore conclude that the data show no statistical detection of the 21-cm forest signal.

As our primary statistical observable, we use the one-dimensional power spectrum of the normalized 21-cm forest transmission, $F_{21}$. Following \citet{Soltinsky_2025}, we define the flux fluctuation estimator as
\begin{equation}\label{eq:PS_estimator}
\delta_{F_{21}} = F_{21} - 1 = \frac{S}{S_{\mathrm{cont}}}-1,
\end{equation}
and compute the 1D power spectrum $P_{21}(\kappa_{\nu})$ from the Fourier transform of $\delta_{F_{21}}$,
\begin{equation}
P_{21}(\kappa_{\nu})\,\delta_D(\kappa_{\nu}-\kappa_{\nu}') = \langle \tilde{\delta}_{F_{21}}(\kappa_{\nu})\,\tilde{\delta}_{F_{21}}^*(\kappa_{\nu}') \rangle,
\end{equation}
where $\delta_D$ is the Dirac delta function. For uniformly sampled spectra with $n$ frequency channels of width $\Delta\nu$, the power spectrum can be estimated using a discrete Fourier transform as
\begin{equation}
P_{21}(\kappa_{\nu,q}) = \left(\frac{2\pi}{n\,\Delta\nu}\right)\left\langle \left|\tilde{\delta}_{F_{21}}(\kappa_{\nu,q})\right|^2 \right\rangle,
\end{equation}
with $q = 0, 1, ..., n-1$ and $k_q=2\pi q/n\,\Delta\nu$ \citep{Soltinsky_2025}. Here the angle brackets denote an average over the Fourier modes contributing to a given $\kappa_{\nu}$-bin.

In practice, the observed spectra contain frequency channels that are removed during the RFI excision and quality-control procedures. Rather than interpolating across these gaps, we estimate the power spectrum using a Lomb--Scargle periodogram, which naturally accommodates irregularly sampled data \citep{Lomb_1976,Scargle_1982,Zechmeister_2009}. The resulting periodogram is normalized to approximately reproduce the convention adopted for uniformly sampled Fourier-transform-based estimates. The effect of missing channels due to the flagging procedure on the measured 21-cm forest 1D power spectrum is discussed in App.~\ref{app:flagging}.

\begin{figure*}
    \begin{minipage}{\linewidth}
 	  \centering
 	  \includegraphics[width=\linewidth]{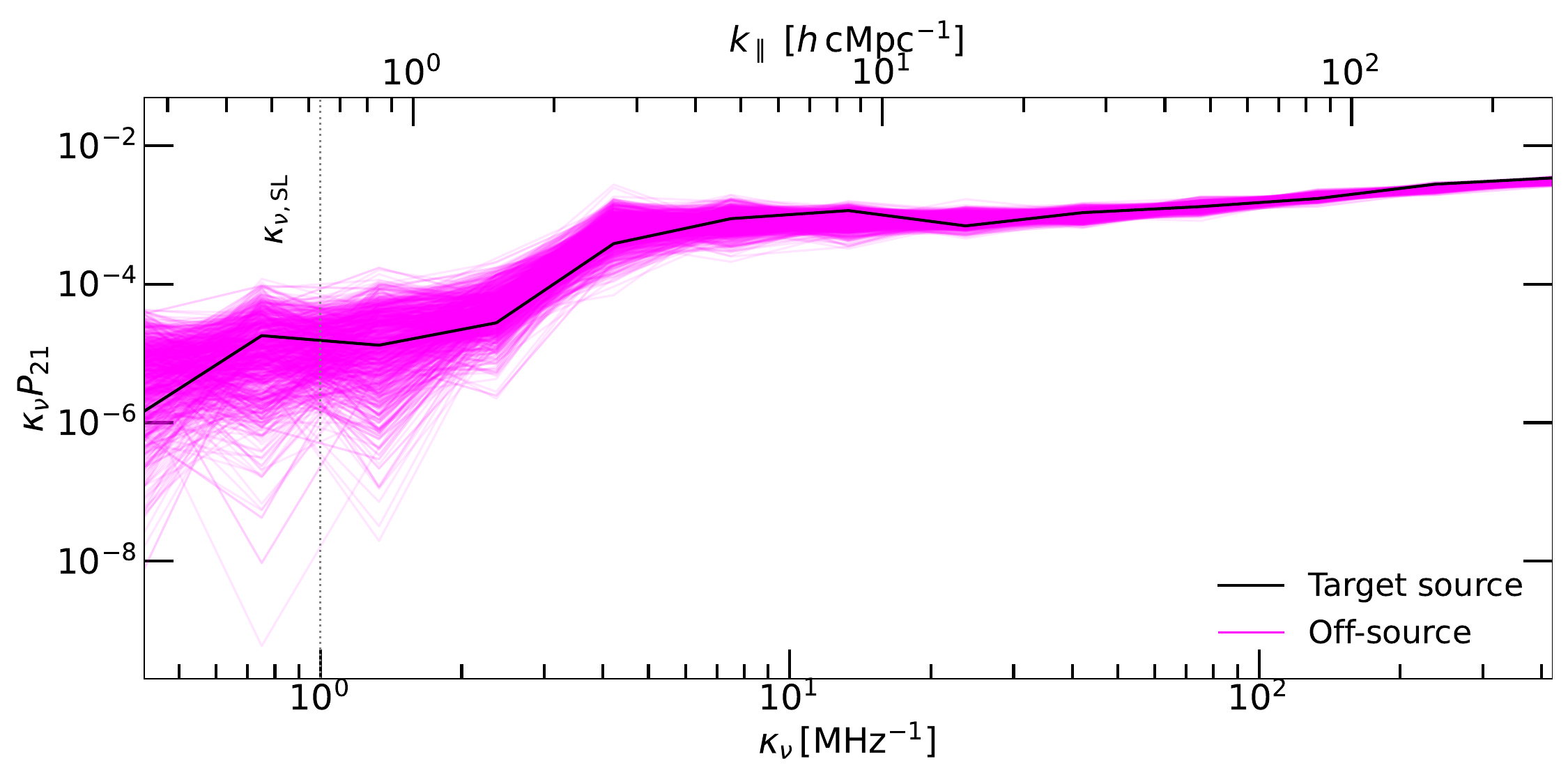}
	\end{minipage}
    \vspace{-0.3cm}
    \caption{Measured $\kappa_{\nu}P_{21}$ (black curve) from the J352--15 spectrum (Fig.~\ref{fig:F21_obs}). For comparison the pink curves mark power spectra of 1000 randomly selected off-source spectra close to the target source (grey curves in Fig.~\ref{fig:rms}) corresponding to the instrumental noise. The highest $\kappa_{\nu}$-bin expected to be dominated by sidelobe contamination is marked with the grey vertical dotted line.}
    \label{fig:1DPS_obs}
\end{figure*}

Throughout the analysis we work with the dimensionless quantity $\kappa_{\nu}P_{21}$ evaluated in 13 logarithmically spaced (0.25\,dex) bins spanning $k \in [0.316,\,562.341]\,\mathrm{MHz^{-1}}$ as presented in Fig.~\ref{fig:1DPS_obs}. Here the black curve corresponds to the measured $P_{21}$ of the target spectrum and represents, to our knowledge, the first observational measurement of the 1D power spectrum of the 21-cm forest during the Epoch of Reionization.

Similarly, we present the 1D power spectrum of the 1000 off-source spectra as pink curves. The signal $\kappa_{\nu}P_{21}$ aligns with the spread of the noise-only cases, hence further supporting the null-detection interpretation of the 21-cm forest signal in the uGMRT data.

Note that for the following analysis we exclude the $\kappa_{\nu}<\kappa_{\nu,\rm SL}=1\,\rm MHz^{-1}$ (indicated by the vertical dashed grey line in Fig.~\ref{fig:1DPS_obs}) bins due to the sidelobe-induced mode mixing as described in Sec.~\ref{sec:sidelobes}.

\subsection{Comparison with archival MWA observations}\label{sec:MWA_obs}

The Murchison Widefield Array (MWA) is a low frequency precursor telescope to the SKA-Low, located in outback Western Australia \citep{Bowman_2013_MWA,Tingay_2013,Wayth_2018}. Archival MWA observations of J352--15 obtained between 2013 and 2022 are analysed for comparison, incorporating Phase I and II data, including datasets from projects G0008, G0045, D0016, and D0041. A total of 560 observations (18.7 hours) were identified to meet the observational requirements (observation phase centre within 10 degrees of the J352--15 source, and $190-210\,\rm MHz$ spectral coverage). Data were re-phased to the source RA and Dec, and calibrated with the apparent brightest 8000 sources within 50 degrees, using the Hyperdrive calibration software \citep{MWA_hyperdrive}. Data were then averaged to $40\,\rm kHz$ spectral resolution and 16~second temporal resolution to control file size but retain high spectral resolution. Finally, data were peeled using an ionospheric phase screen model that estimated the offsets of source positions from their catalogue positions. 1000 sources were ionospherically-peeled and the remaining sky model directly subtracted. This subtraction would be expected to reduce the source sidelobes in the data, but not eliminate them. A common set of frequency channels were extracted yielding data at $200.96\,\mathrm{ MHz}\leq\nu\leq231.64\,\rm MHz$.

We employ visibility beamforming to extract the flux density for each spectral channel. Here we combine all of the measured phased visibilities in a weighted sum, and take the real part:
\begin{equation}
    S(\nu) = \text{Re}\frac{\displaystyle\sum_{i=1}^N w_iV_i(\nu)}{\displaystyle\sum_{i=1}^N w_i},
\end{equation}
where $w_i$ is a taper function. In regular imaging, this taper can apply a Briggs weight, or natural or uniform weighting. Here, we apply a Tukey filter with a minimum wavelength of 50$\lambda$, to remove short baselines to avoid diffuse emission. In general, diffuse emission is sub-dominant in MWA data beyond 30--40$\lambda$. The beamformed visibilities and their weights are extracted for each spectral channel for each 2-min MWA snapshot observation, thereby producing a small dataset for further analysis. Spectra were extracted for $z < 5.84$, yielding 602 spectral channels. Additional RFI flagging was performed on each observation, removing those with channels yielding large variances relative to the mean. This produced a final set of 311 observations (9.7 hours) for further analysis.

\begin{figure}
    \begin{minipage}{\linewidth}
 	  \centering
 	  \includegraphics[width=\linewidth]{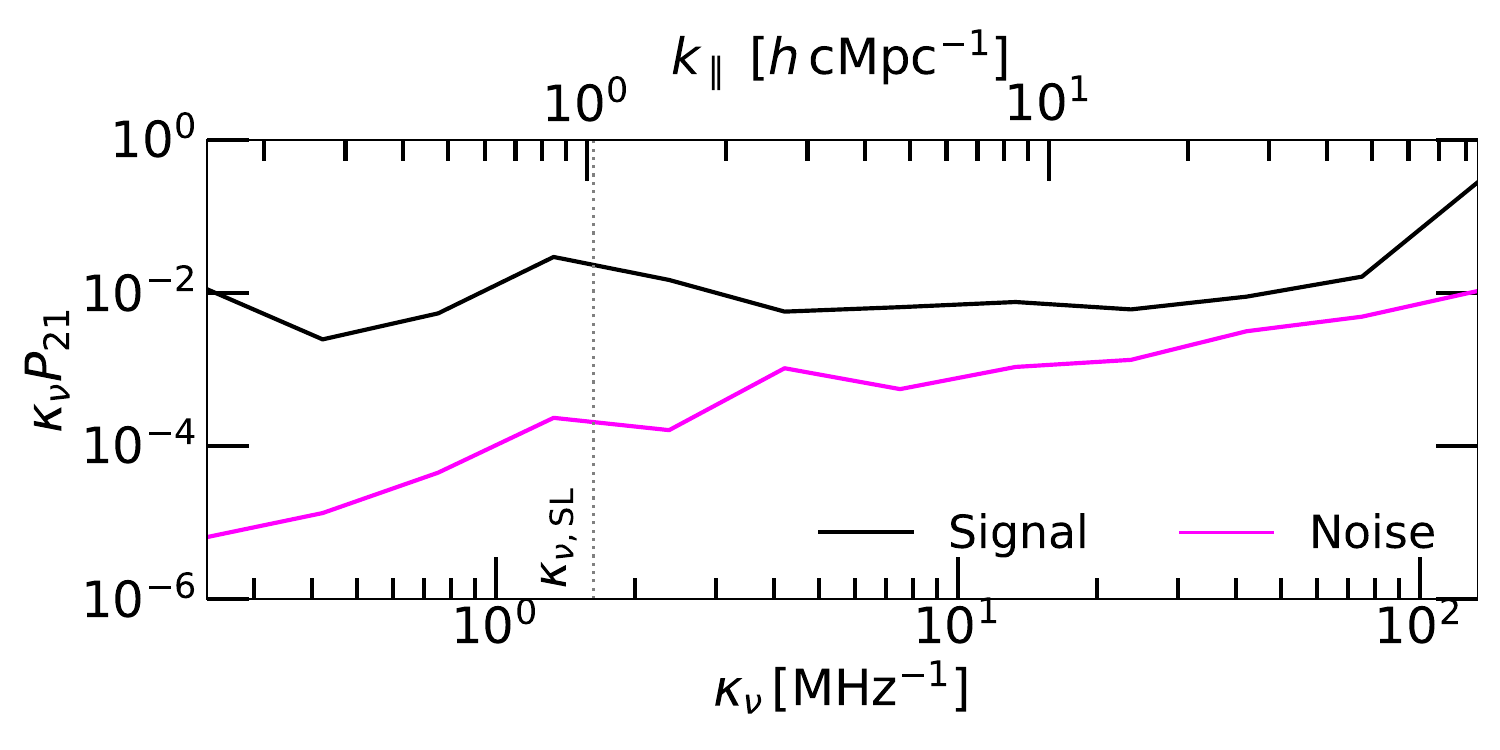}
	\end{minipage}
    \vspace{-0.3cm}
    \caption{Measured $\kappa_{\nu}P_{21}$ from the MWA observations of J352--15 (black) compared to the empirically estimated noise power spectrum (pink). The vertical dotted grey line indicates the $\kappa_{\nu}$-bin up to which the sidelobe contamination is expected to dominate.}
    \label{fig:1DPS_MWA}
\end{figure}

As described in Sec.~\ref{sec:sidelobes}, unlike uGMRT, MWA's large primary beam yields a larger number of modes that should be theoretically sidelobe-limited. For these data, that limit is $\kappa_{\nu,SL} = 1.6\,\rm MHz^{-1}$. MWA also has lower sensitivity than uGMRT, with a theoretical noise level on $40\,\rm kHz$ channels of $120\,\rm mJy\,beam^{-1}$ for 9.7 hours of data. The remaining analysis matches that for the uGMRT data. Figure \ref{fig:1DPS_MWA} shows the final power spectrum (black) and an estimate of the noise (pink). The noise estimate is empirically determined from time-differenced visibilities, yielding a difference spectrum noise level of $100\,\rm mJy\,beam^{-1}$, which is close to that predicted by the radiometer equation ($120\,\rm mJy\,beam^{-1}$). At high $\kappa_{\nu}$, the measured spectrum shows noise-like behaviour. The limits are significantly poorer than for the uGMRT, given the large FoV, and the lower sensitivity, but show promise for further investigations with MWA. Consequently, we do not incorporate the MWA measurements into the inference analysis presented below.

\section{Forward modelling the 21-cm forest signal}\label{sec:mocks}

While it is possible to model the 21-cm forest 1D power spectrum analytically \citep{Shao_2025_analytical}, we choose a semi-numerical approach which allows us to explore many thermal and ionization models of the IGM. Our modelling of the IGM and the forward construction of synthetic 21-cm forest spectra follows the framework developed in \citet{Soltinsky_2025} and subsequently adopted in \citet{Patil_2026}. Here we summarise only the key elements relevant for the present analysis and refer the reader to those works for a detailed description.

\subsection{Modelling IGM during reionization}

We use the models of the IGM at $z=6$ generated in \citet{Soltinsky_2025}. These are based on the semi-numerical code \texttt{21cmFAST}\footnote{\hyperlink{https://github.com/21cmfast/21cmFAST}{https://github.com/21cmfast/21cmFAST}} version 3.3.1 \citep{Mesinger_2011_21CMFAST, Murray_2020}, assuming a flat $\Lambda$CDM cosmology consistent with Planck measurements, specifically $\Omega_{\Lambda}=0.692$, $\Omega_{\rm m}=0.308$, $\Omega_{\rm b}=0.0482$, $\sigma_8=0.829$, $n_{\rm s} =0.961$ and $h=0.678$ \citep{planck2014}. The simulations are performed in a $(50\,\mathrm{cMpc})^3$ volume with $256^3$ cells.

The ionization state of the IGM is characterised by the mean neutral hydrogen fraction $\langle x_{\mathrm{HI}} \rangle$, which is controlled by varying the ionizing efficiency of galaxies (simulation parameter \textsc{HII\_EFF\_FACTOR}) while fixing the mean free path of ionizing photons (simulation parameter \textsc{R\_BUBBLE\_MAX}) to $0.75\,\rm pMpc$, consistent with measurements at $z\sim6$ \citep{Becker_2021}. The thermal state of the neutral IGM is governed by the X-ray background radiation efficiency $f_{\rm X}$, which sets the X-ray luminosity per unit star-formation rate, $SFR$, (simulation parameter \textsc{L\_X}) following \citet{Furlanetto_2006b}, where
\begin{equation}
L_{\rm X}=3.4\times 10^{40}\rm\,erg\,s^{-1} \, \mathnormal{f_{\rm X}} \left(\frac{SFR}{1\,M_{\odot}\rm\,yr^{-1}}\right). \label{eq:LX} 
\end{equation}
\noindent
The simulations span the physically relevant ranges $\langle x_{\rm HI}\rangle=[0,1]$ and $\log_{10}f_{\rm X}=[-4,1]$ over 534 IGM models distributed across the parameter space shown in Fig.~2 of \citet{Patil_2026}. Although the simulations are generated by varying the ionizing efficiency and X-ray efficiency parameters, we express our results in terms of the derived quantities $\langle x_{\rm HI}\rangle$ and $\langle T_{\rm HI}\rangle$ to make the inference more directly connected to the physical state of the IGM. We define $\langle T_{\rm HI}\rangle$ as the volume-averaged temperature of predominantly neutral cells with $x_{\rm HI}\geq0.9$. In our simulations, $\langle T_{\rm HI}\rangle$ ranges from $\sim14\,\mathrm{K}$ to $\sim10000\,\mathrm{K}$, with the highest values corresponding to models with very efficient X-ray heating prior to the completion of reionization.


We note the slight mismatch between the simulation redshift ($z=6$) and the quasar redshift ($z=5.82$). Since our inference is expressed in terms of the physical IGM quantities $\langle x_{\rm HI}\rangle$ and $\langle T_{\rm HI}\rangle$, we do not expect this modest redshift offset to significantly affect our conclusions\footnote{The corresponding change in the mean cosmic density is $\sim8\%$ assuming $\rho\propto(1+z)^3$, while the cosmic time difference is only $\sim40\,\rm Myr$, both small compared to the timescales over which the thermal and ionization state of the IGM evolve during the late stages of reionization.}.

The transverse cell size of the simulations is $\sim195\,\mathrm{ckpc}$. Since J352--15 is unresolved in our observations, the synthesized beam provides only an upper limit on the intrinsic angular extent of the radio-emitting region. We approximate the background radio emission as a point source and model the 21-cm forest using individual LOS through the simulations. From each simulation we extract 1000 LOS skewers distributed approximately equally among the three Cartesian directions and uniformly spaced in the transverse plane, providing a regular sampling of the simulation volume. Along each LOS we sample gas overdensity, $\Delta$, peculiar velocity, $v_{\rm pec}$, neutral fraction, $x_{\mathrm{HI}}$, and kinetic temperature of the gas, $T_{\rm K}$, which form the basis for constructing synthetic 21-cm forest spectra. We assume a saturated coupling between the spin temperature, $T_{\rm S}$, and $T_{\rm K}$. \citet{Soltinsky_2021} show that the difference in 21-cm forest absorption features distribution between saturated and unsaturated Wouthuysen-Field coupling scenarios is negligible at $z=6$, and hence $T_{\rm S}=T_{\rm K}$ is a reasonable assumption here.

\subsection{Forward model for the 21 cm forest 1D power spectrum}

The normalized 21-cm forest flux, besides as defined in Eq.~\ref{eq:PS_estimator}, is also given by $F_{21} = e^{-\tau_{21}}$, where the discrete form of the 21-cm forest optical depth is computed along each LOS following the standard formalism \citep{Furlanetto_Loeb_2002}
\begin{align}
\tau_{\rm 21, i} =~& \frac{3h_{\rm p}c^{3}A_{10} }{32\pi^{3/2}\nu_{21}^{2}k_{\rm B}} \frac{\delta v}{H(z)} \nonumber \\
             & \times \sum_{j=1}^{N}\frac{\Delta_{j}x_{{\rm HI}, j}}{b_{j}T_{{\rm S}, j}}\exp\left( - \frac{ (v_{{\rm H},i}-u_{j})^{2}} {b^{2}_{j}}\right), \label{eq:tau21_discrete}
\end{align}
\noindent
where the Doppler parameter $b=(2k_{\rm B}T_{\rm K}/m_{\rm H})^{1/2}$, $h_{\rm p}$ and $k_{\rm B}$ are the Planck and Boltzmann constants, respectively, $c$ is the speed of light, $A_{10}=2.85\times 10^{-15}\rm\,s^{-1}$ is the Einstein spontaneous emission coefficient for the hyperfine transition. To achieve properly converged spectra over line profile we resample the LOS using linear interpolation such that the velocity width of the synthetic pixels is smaller than $b$. Individual spectra are randomly spliced to produce bandwidths comparable to the observational setup.

\begin{figure}
    \begin{minipage}{\linewidth}
 	  \centering
 	  \includegraphics[width=\linewidth]{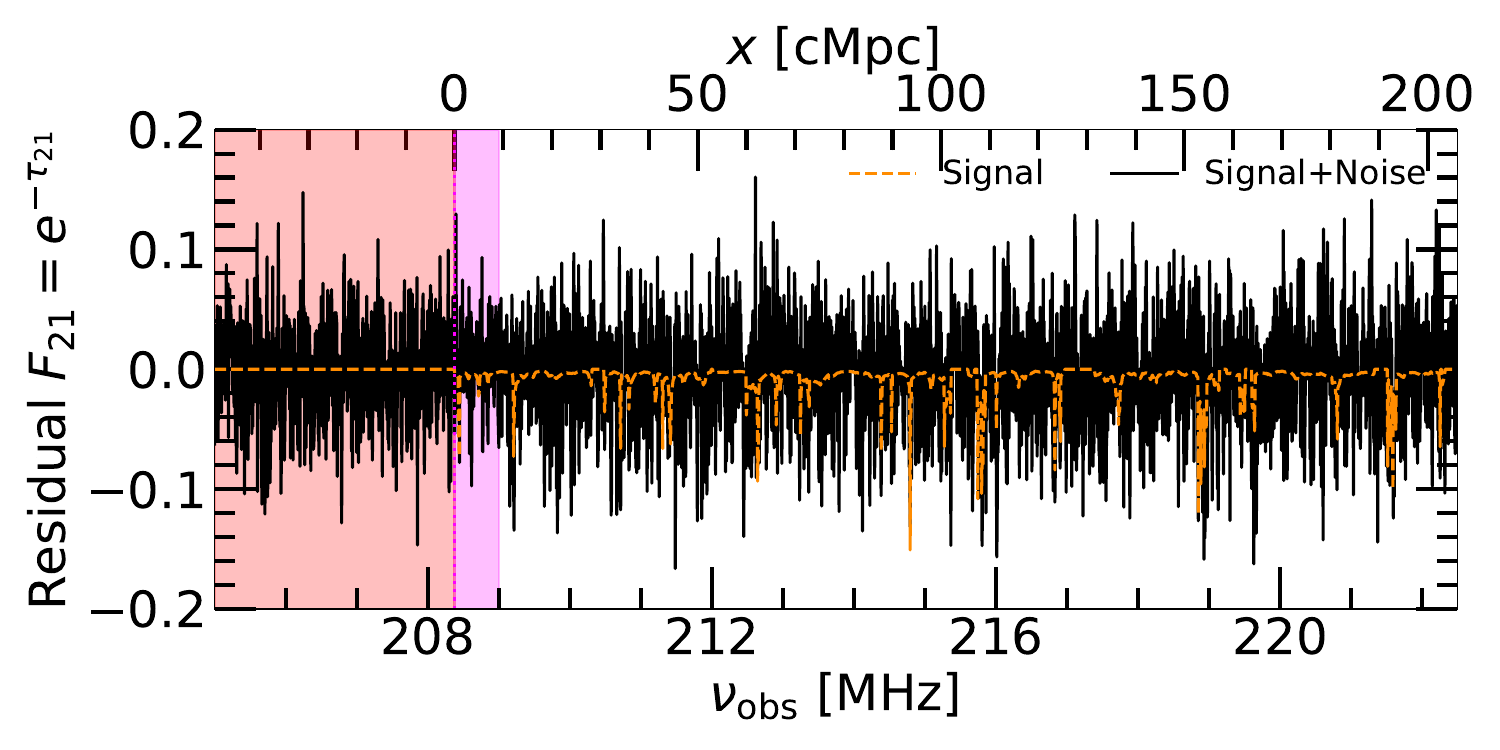}      
 	  \includegraphics[width=\linewidth]{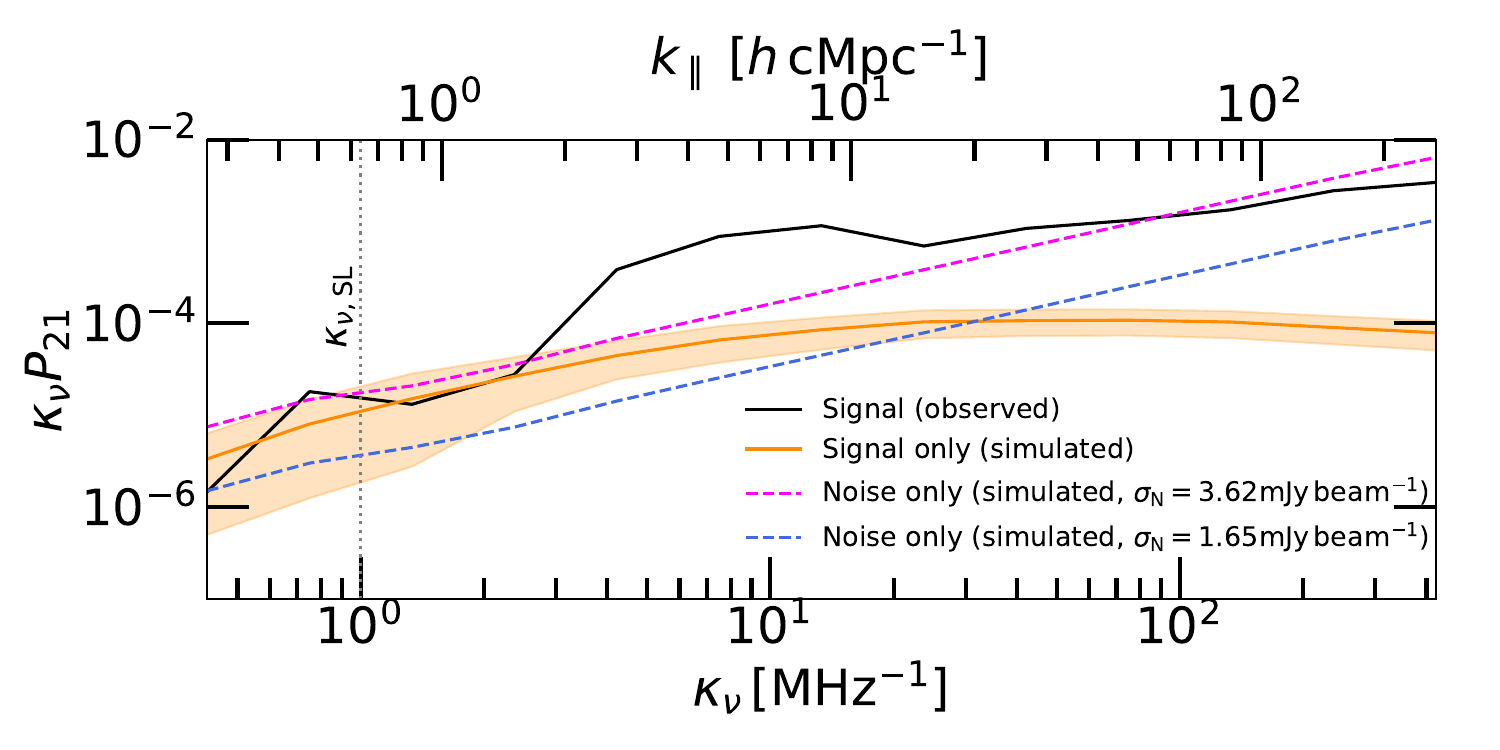}
	\end{minipage}
    \vspace{-0.3cm}
    \caption{\textit{Top:} A synthetic spectrum of J352--15 at $z \simeq 5.82$, normalised by the continuum, shown both without thermal noise (orange dashed) and with thermal noise (black solid) for a total of $t_{\rm eff}=11.73\,\mathrm{hr}$, corresponding to $\sigma_{\mathrm{N}}=3.62\,\mathrm{mJy\,beam^{-1}}$, in an IGM model with $\langle x_{\rm HI}\rangle=0.89$ and $\langle T_{\rm HI}\rangle=16.2\,\rm K$. The pink vertical dotted line marks the rest-frame 21-cm frequency; 21-cm forest absorption is expected only bluewards. Hence, the red shaded region marks the frequencies not used in the analysis. The pink shaded region indicates $208.4-209.0\,\mathrm{MHz}$ range that is excluded from our analysis due to the potential \HI~absorption associated with the host galaxy and the proximity zones. \textit{Bottom:} Simulated 1D forest power spectrum, $\kappa_{\nu}P_{21}$ (orange) for the same IGM model; solid curve marks the mean over 1000 LOS while shading indicates $68\%$ scatter. Dashed curves show noise limits for $\sigma_{\mathrm{N}}=3.62\,\mathrm{mJy\,beam^{-1}}$ (pink; matching the top panel) and $1.65\,\mathrm{mJy\,beam^{-1}}$ (blue). We show the measured $\kappa_{\nu}P_{21}$ (solid black curve) for comparison. The grey vertical dotted line indicates the highest $\kappa_{\nu}$-bin expected to be dominated by sidelobe contamination.}
    \label{fig:1DPS_sim}
\end{figure}

The synthetic spectra are convolved with a boxcar kernel to match the observational spectral resolution of $\delta\nu=6.1\,\rm kHz$. For the illustrative spectrum shown in Fig.~\ref{fig:1DPS_sim}, we add Gaussian thermal noise assuming frequency-independent $\sigma_{\rm N}=3.62\,\rm mJy\,beam^{-1}$ which is then normalized by the intrinsic flux density spectrum. An example simulated flux spectrum in a rather extreme IGM model with cold ($\langle T_{\rm HI}\rangle=16.2\,\rm K$) and neutral gas ($\langle x_{\rm HI}\rangle=0.89$) is presented in the top panel of Fig.~\ref{fig:1DPS_sim} shown both without thermal noise (orange dashed curve) and after the addition of thermal noise (solid black curve). As explained in Sec.~\ref{sec:target_selection}, due to a potential \HI~absorption associated with the host galaxy and background quasar radiation induced effects on the surrounding gas, we consider only $\nu\geq209\,\mathrm{MHz}$ in our analysis (i.e. red and pink shaded regions in Fig.~\ref{fig:1DPS_sim} is excluded).

Fig.~\ref{fig:1DPS_sim} also shows the synthetic 1D power spectra of the 21-cm forest signal only, $\kappa_{\nu}P_{21}^{\rm S}$ where the solid orange curve is the mean over 1000 synthetic LOS and the orange shaded region is $68\%$ scatter around it. This is computed considering the same IGM model as in the top panel. The noise limit, $\langle \kappa_{\nu}P_{21}^{\rm N}\rangle$, is computed as the mean thermal-noise-only power spectrum estimated from 1000 white-Gaussian-noise realizations with variance $\sigma_{\rm N}=3.62\,\rm mJy\,beam^{-1}$ marked by the dashed pink curve. One can see that even in this scenario, that only a minority of signal realizations approach the noise level.

At $\kappa_{\nu}\lesssim3\,\mathrm{MHz^{-1}}$ both simulated $\kappa_{\nu}P_{21}^{\rm S}$ (solid orange curve) and $\kappa_{\nu}P_{21}^{\rm N}$ (dashed pink curve) broadly agree with the measured power spectrum (solid black curve) in Fig.~\ref{fig:1DPS_obs} with amplitude of $\sim10^{-6}-10^{-5}$. At smaller scales the measured $\kappa_{\nu}P_{21}$ aligns better with the pink curve rather than the orange one, as one would expect from noise-dominated measurement. The overall scale dependence is qualitatively similar with the dimensionless power spectra reaching $\kappa_{\nu}P_{21}\sim10^{-3}$ at $\kappa_{\nu}\sim500\,\mathrm{MHz^{-1}}$. However, the observed spectrum exhibits excess power at intermediate scales, $3\,\mathrm{MHz^{-1}}\lesssim \kappa_{\nu} \lesssim 20\,\mathrm{MHz^{-1}}$, likely reflecting residual systematic effects beyond idealized thermal noise.

\begin{figure*}
    \begin{minipage}{0.33\linewidth}
 	  \centering
 	  \includegraphics[width=\linewidth]{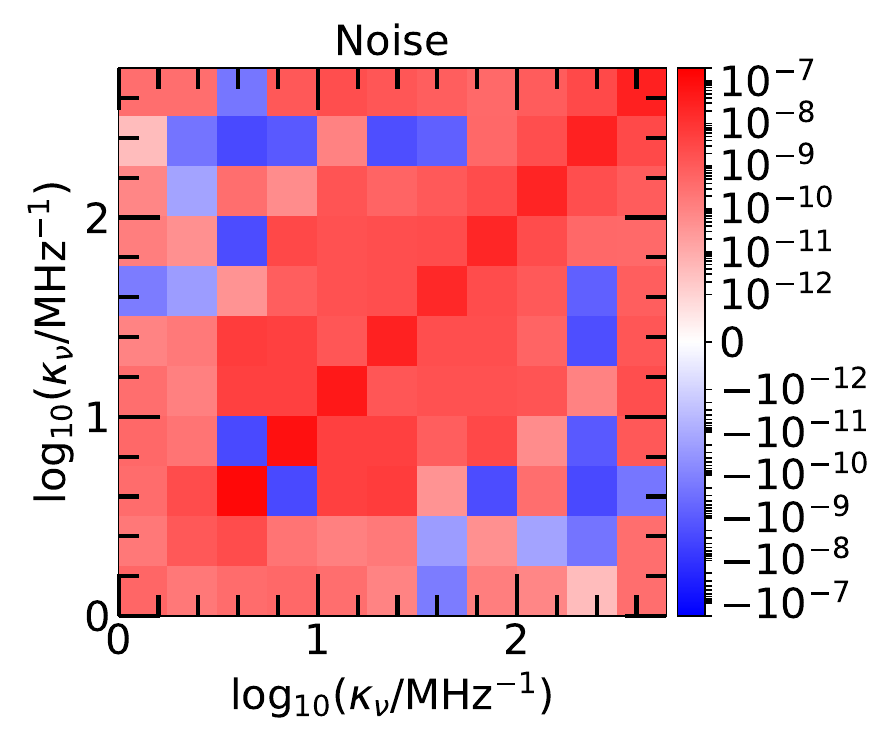}
	\end{minipage}
     \begin{minipage}{0.33\linewidth}
 	  \centering
 	  \includegraphics[width=\linewidth]{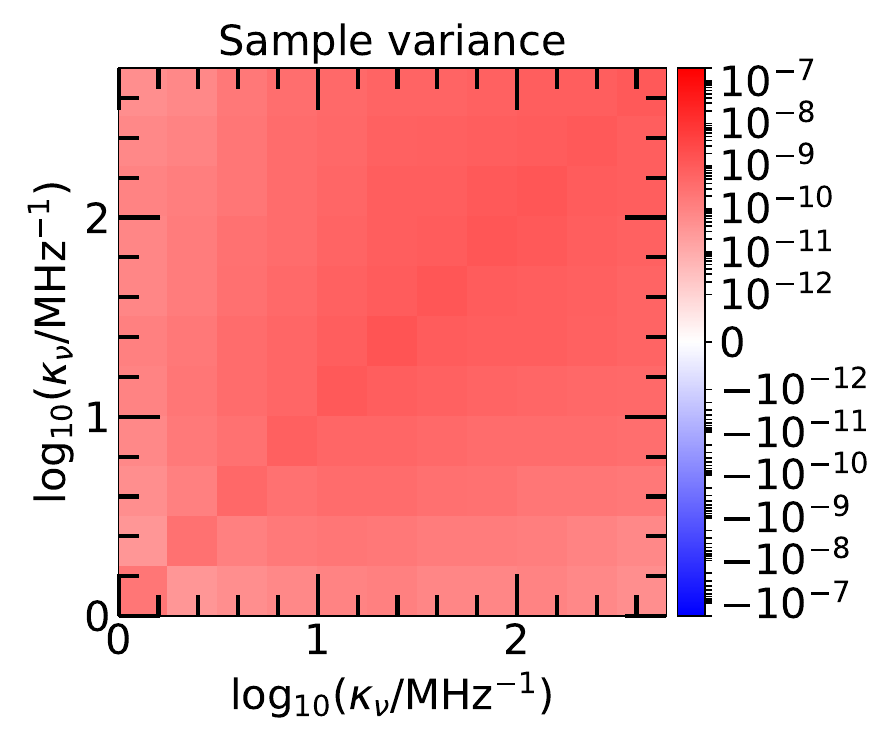}
	\end{minipage}
      \begin{minipage}{0.33\linewidth}
 	  \centering
 	  \includegraphics[width=\linewidth]{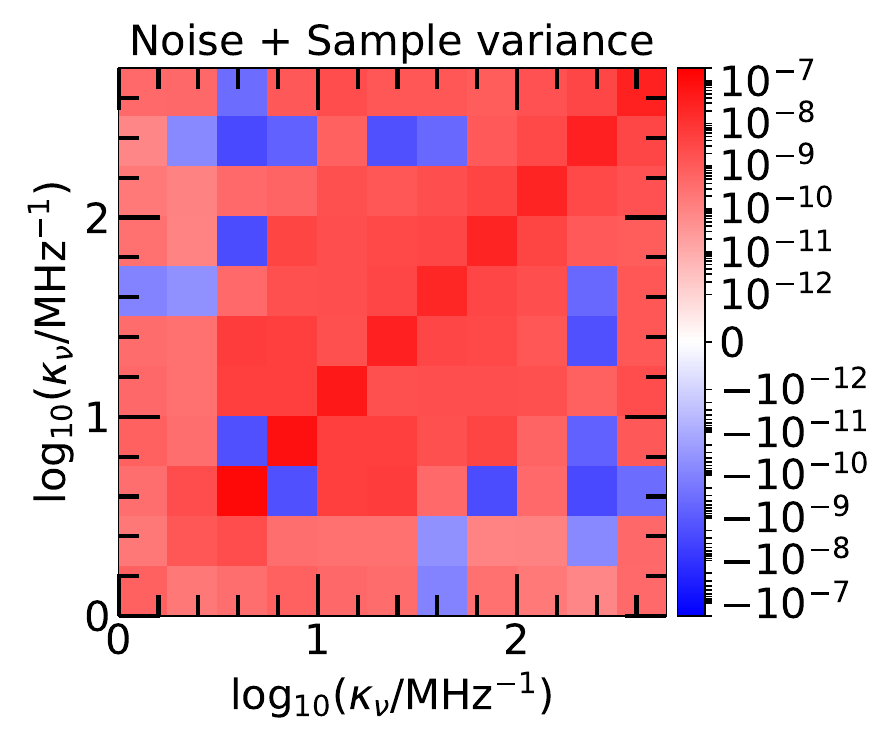}
	\end{minipage}
	\vspace{-0.3cm}
    \caption{Covariance matrix for the noise variance (measured, left panel), sample variance (simulated assuming an IGM model of $\langle x_{\rm HI}\rangle=0.89$ and $\langle T_{\rm HI}\rangle=16.2\,\rm K$, middle panel) and their combined effect on the observations (right panel).}
    \label{fig:covariance_matrices}
\end{figure*}

\section{IGM constraints from the J352--15 1D power spectrum}\label{sec:constraints}

As shown in \citet{Soltinsky_2025}, even a null-detection of the 21-cm forest 1D power spectrum can give meaningful constraints on the thermal and ionization state of the IGM. Here we adopt the method A2 from \citet{Patil_2026}\footnote{We do not employ the AI/ML based methods in \citet{Patil_2026} because they have not been tested for a null-detection case as opposed to method A2.}. Specifically, we run Markov Chain Monte Carlo analysis \citep[MCMC,][]{Goodman_2010}\footnote{Our MCMC analysis is initiated with 64 walkers for each inferred parameter using \textsc{emcee} \citep{Foreman-Mackey_2013_emcee} and \textsc{corner} \citep{ForemanMackey_2016_corner} software packages. The chains start at initial guesses uniformly spread over the whole priors and consist of $10^6$ steps with first 1000 then excluded.} assuming a Gaussian likelihood 
\begin{equation}
    \mathcal{L}(P_{21}\lvert\boldsymbol{\theta})=\frac{1}{\sqrt{(2\pi)^{N_{\rm bins}}\mathrm{det\,\textbf{C}(\boldsymbol{\theta})}}}\exp\left(-\frac{1}{2}\mathrm{\mathbf{r}(\boldsymbol{\theta})^T\textbf{C}(\boldsymbol{\theta})^{-1}\mathbf{r}(\boldsymbol{\theta})}\right),\label{eq:likelihood_covar}
\end{equation}
where $\mathbf{r}(\boldsymbol{\theta})$ is the residual vector, the parameter vector is $\boldsymbol{\theta}=\{\log_{10}(\langle T_{\mathrm{HI}}\rangle/\mathrm{K}),\langle x_{\mathrm{HI}}\rangle\}$, $\mathbf{C}(\boldsymbol{\theta})$ is the model-dependent covariance matrix, and $N_{\rm bins}=11$ given that we omit the two $\kappa_{\nu}$-bins at $<1\,\rm MHz^{-1}$.

Here we compare measured noise-subtracted power spectrum, $P_{21}^{\rm N_{\rm sub}}$, to the simulated signal-only power spectrum via
\begin{equation}
\mathbf{r}(\boldsymbol{\theta})=P_{21}^{\rm N_{\rm sub}}-P_{21}^{\rm sim}(\boldsymbol{\theta})=\left(P_{21}^{\rm obs}-\langle P_{21}^{\rm N}\rangle\right)-P_{21}^{\rm sim}(\boldsymbol{\theta}).
\end{equation}
This method exploits the fact that we have pure noise measurements because of the availability of off-target sight lines in our data. Note that while \citet{Patil_2026} showed that the 21-cm forest is not fully Gaussian (see their appendix A), the Gaussian likelihood provides robust results.

Uniform prior on $\langle x_{\mathrm{HI}}\rangle$ spanning the whole physically possible range from 0 to 1 is assumed. For $\langle T_{\rm HI}\rangle$, we restrict the analysis to the range $15\,\mathrm{K}$--$10^4\,\mathrm{K}$. While the lower value is limited by the simulations\footnote{Note that not all of the $\langle x_{\mathrm{HI}}\rangle$ cases in our IGM models reach down to $\langle T_{\mathrm{HI}}\rangle=14\,\mathrm{K}$, hence we select a slightly higher lower bound on the prior. 
}, the upper bound corresponds to the temperatures reached in reionization bubbles where no 21-cm forest signal would be detected\footnote{Note that \citet{Soltinsky_2021} found that 21-cm forest features with $F_{21}\lesssim0.99$ require IGM temperatures of $\lesssim100\,\mathrm{K}$.}. Given the orders-of-magnitude range of neutral IGM temperatures, we implement a uniform prior in $\log_{10}(\langle T_{\mathrm{HI}}\rangle/\mathrm{K})=[\log_{10}(15),4]$. This choice of $\boldsymbol{\theta}$ allows the inferred constraints to be interpreted directly in terms of the physical thermal state of the neutral IGM, independent of the specific heating model used to generate the simulations.

We now use the measured 1D power spectrum from Sec.~\ref{sec:obs_spectrum} as the observational input to the inference framework. To our knowledge, this is the first application of a 21-cm forest 1D power spectrum inference framework to observational data. The mean noise-only power spectrum, $\langle P_{21}^{\rm N}\rangle$, is estimated by averaging the power spectra of all off-source sightlines (pink curves in Fig.~\ref{fig:1DPS_obs}).

These are also used to compute the covariance matrix, $\textbf{C}(\boldsymbol{\theta})$, presented in Fig.~\ref{fig:covariance_matrices}. In the left panel, the $\textbf{C}(\boldsymbol{\theta})$ representing the uncertainty from the instrumental noise only is shown. We estimate the contribution of the sample variance by generating 1000 realizations of $P_{21}^{\rm sim}(\boldsymbol{\theta})$ as described in Sec.~\ref{sec:mocks}. An example of $\langle x_{\rm HI}\rangle=0.89$ and $\langle T_{\rm HI}\rangle=16.2\,\rm K$ IGM model (orange shaded band in Fig.~\ref{fig:1DPS_sim}) is shown in the middle panel. In this case, the sample variance contribution is significantly smaller than the instrumental-noise contribution, with diagonal covariance elements that are typically a factor of $\sim2-200$ lower. 
This contrasts with the forecasts of \citet{Soltinsky_2025}, where substantially longer integration times and idealized noise computation based on radiometer equation were assumed, resulting in much lower thermal-noise levels and a correspondingly larger relative contribution from sample variance. Given that we incorporate this effect, the covariance matrix depends on the parameter values tested in the MCMC step when computing $\mathcal{L}(P_{21}\lvert\boldsymbol{\theta})$. Finally, we combine the effect of noise and sample variance by computing the covariance matrix from an ensemble of $P_{21}^{\rm N}+P_{21}^{\rm sim}(\boldsymbol{\theta})$. This is shown in the right panel of Fig.~\ref{fig:covariance_matrices}.

\begin{figure*}
    \begin{minipage}{\linewidth}
 	  \centering
 	  \includegraphics[width=0.7\linewidth]{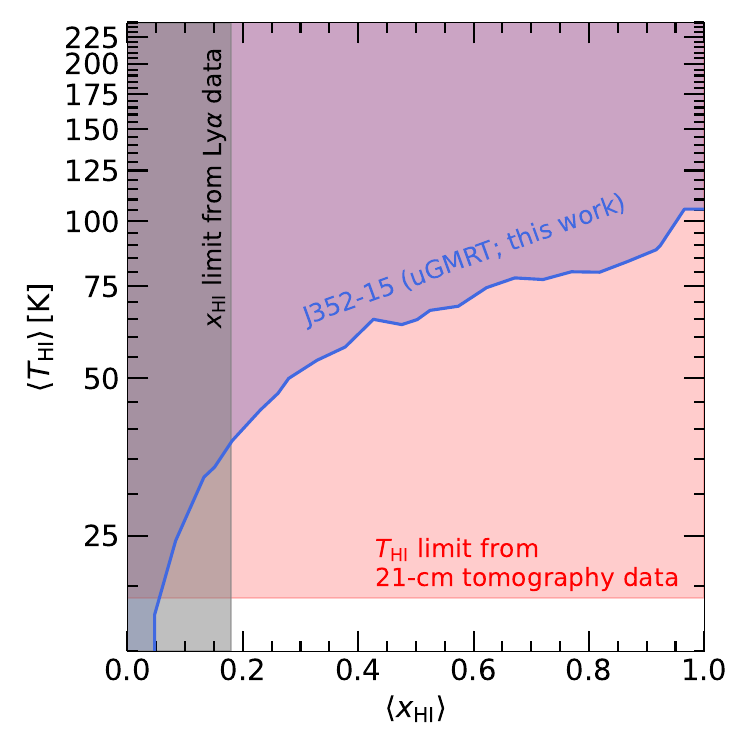}      
	\end{minipage}
    \vspace{-0.3cm}
    \caption{Constraints on the $\langle T_{\mathrm{HI}}\rangle$--$\langle x_{\rm HI}\rangle$ plane based on a \emph{null detection} of the 21-cm forest 1D power spectrum from the archival observations of $t_{\rm eff}=11.73\,\mathrm{hr}$ and the rms noise level of $\sigma_{\rm N}=3.62\,\mathrm{mJy\,beam^{-1}}$ targeting J352--15, a RLQSO at $z=5.82$. The blue shaded region indicates the $68\%$ credible interval favoured parameter space. For comparison, the grey vertical band indicates current Ly$\alpha$-based constraints on $\langle x_{\rm HI}\rangle$ \citep{Gaikwad_2023,Umeda_2025,Qin_2025}, while the red horizontal band shows existing limits on the IGM spin temperature from the cosmic X-ray background, UV luminosity functions, and 21-cm tomography measurements \citep{Dhandha_2025}.}
    \label{fig:null-limits_archival}
\end{figure*}

Applying the inference framework described above to the measured 1D power spectrum yields the constraints shown in Fig.~\ref{fig:null-limits_archival}. We provide the whole corner plot including marginalized posterior distributions of both IGM parameters in App.~\ref{app:corner_plot}. The Bayesian analysis yields a broad posterior distribution (blue shaded region). Rather than identifying a preferred IGM model, the data primarily exclude regions of the $\langle T_{\rm HI}\rangle$--$\langle x_{\rm HI}\rangle$ parameter space. In particular, models lying below and to the right of the blue region are disfavoured at the $68\%$ credible level.

It is important to note that these constraints are not obtained by asking whether the predicted 21-cm forest power spectrum exceeds the mean noise power spectrum shown in Fig.~\ref{fig:1DPS_sim}. Instead, the likelihood compares the measured residual power spectrum, $P_{\rm obs}-\langle P_{\rm N}\rangle$, with the model prediction, $P_{21}^{\rm sim}(\boldsymbol{\theta})$, using the full covariance matrix. Consequently, the relevant quantity is the uncertainty on the residual power spectrum after subtraction of the estimated noise bias rather than the amplitude of the mean noise power spectrum itself. Models can therefore be disfavoured even when their predicted power spectrum lies below the raw noise level in individual $\kappa_{\nu}$-bins, as the likelihood combines information from all retained $\kappa_{\nu}$-modes simultaneously and accounts for their covariance.

The overall trend of the exclusion boundary reflects the approximate scaling of the 21-cm optical depth with $x_{\rm HI}/T_{\rm HI}$ (see Eq.~\ref{eq:tau21_discrete}). Colder and more neutral IGM models produce stronger absorption fluctuations and hence a larger 1D power spectrum, resulting in larger residuals and making them increasingly inconsistent with the measured data. Consequently, increasingly higher temperatures are required to remain consistent with the observations as the neutral fraction increases.

These exclusions can be directly interpreted as a joint upper limit on the volume-averaged neutral fraction, $\langle x_{\rm HI}\rangle$, and a lower limit on the mean temperature of the predominantly neutral IGM, $\langle T_{\rm HI}\rangle$. The latter is particularly valuable because the thermal state of neutral gas at $z\gtrsim5.5$ remains largely inaccessible to existing observational probes. For example, our measurement of the 21-cm forest 1D power spectrum disfavours models with $\langle T_{\rm HI}\rangle<[39.8,64.7,79.9]\,\mathrm{K}$ at $\langle x_{\rm HI}\rangle=[0.2,0.5,0.8]$ and $\langle x_{\rm HI}\rangle\gtrsim0.05$ for the coldest IGM models ($\langle T_{\rm HI}\rangle\lesssim18\,\mathrm{K}$) with $68\%$ credible level. Similarly, in the regime allowed by the currently available Ly$\alpha$ observations, $\langle T_{\rm HI}\rangle<26.9\,\mathrm{K}$ at $\langle x_{\rm HI}\rangle=0.1$ models are excluded.

Fig.~\ref{fig:null-limits_archival} also illustrates the complementarity of the 21-cm forest with current Ly$\alpha$-based probes \citep[$68\%$ C.I.][]{Gaikwad_2023,Umeda_2025,Qin_2025} and limits derived from global-signal (SARAS 3) and tomographic 21-cm (HERA) observations \citep[$95\%$ C.I.][]{Dhandha_2025}. The Ly$\alpha$ constraints shown in the figure correspond to measurements at similar redshifts, $5.5\lesssim z\lesssim5.75$, as probed by our observations ($\nu=210-220\,\rm MHz$). By contrast, the temperature constraint from \citet{Dhandha_2025} is quoted at $z=6$, owing to the lack of comparable measurements at lower redshifts.

These probes constrain the IGM in fundamentally different ways. Ly$\alpha$ forest observations are highly sensitive to trace amounts of neutral hydrogen and therefore provide powerful constraints on the ionization state of the IGM. However, because Ly$\alpha$ absorption rapidly saturates in substantially neutral regions \citep[e.g.][]{Fan_2006}, these observations provide little direct information on the temperature of predominantly neutral gas. Conversely, current global-signal and tomographic 21-cm experiments are directly sensitive to the thermal state of neutral hydrogen, but their interpretation relies on separating a faint cosmological signal from foreground emission that is many orders of magnitude brighter. As a result, the corresponding constraints are typically obtained only after marginalising over a large number of astrophysical and instrumental parameters \citep[e.g.][]{Ghara_2025,Dhandha_2025}.

While the constraints presented are relatively broad, they already exclude regions of the $\langle T_{\rm HI}\rangle$--$\langle x_{\rm HI}\rangle$ parameter space that remain allowed by existing observations. This is because the 21-cm forest is simultaneously sensitive to both the abundance of neutral hydrogen and its temperature. As a result, it constrains a joint region of parameter space rather than either quantity individually. Within the adopted modelling framework, the present 21-cm forest power spectrum measurements exclude some of the coldest and most neutral IGM models that remain consistent with existing Ly$\alpha$ and tomographic 21-cm constraints. Furthermore, unlike Ly$\alpha$ and current tomographic 21-cm experiments, which primarily constrain either the ionization or thermal state of the IGM separately, the 21-cm forest provides simultaneous constraints on both quantities. In this sense, even observations of the present sensitivity yield astrophysically informative constraints on the nature of the UV and X-ray sources responsible for reionization and IGM heating.

\section{Forecasts for future observations}\label{sec:constraints_forecasts}

While the archival observations already demonstrate that statistically informative constraints can be obtained from 21-cm forest power-spectrum measurements, their sensitivity remains insufficient to strongly constrain the thermal and ionization state of the IGM. We therefore explore the prospects for future observations by considering a hypothetical $t_{\rm int}=100\,\rm hr$ observing campaign.

As argued in Sec.~\ref{sec:calibration}, we conservatively assume that $60\%$ of data is kept after RFI flagging and calibration quality cuts. A nominal $100\,\rm hr$ observing campaign would therefore provide $t_{\rm eff}=60\,\rm hr$, which combined with the existing archival data yields a total effective integration time of $t_{\rm eff}=71.73\,\rm hr$. Extrapolating the empirical scaling in Eq.~\ref{eq:sigma_N}, we estimate that such a dataset would reach a sensitivity of $\sigma_{\rm N}=1.65\,\rm mJy\,beam^{-1}$ per channel. This forecast assumes that the empirical noise scaling measured from the archival data continues to apply at longer effective integration times. While this assumption is consistent with the available observations, deeper observations may eventually become limited by residual systematics, calibration uncertainties, or persistent RFI, resulting in a weaker sensitivity improvement than predicted by Eq.~\ref{eq:sigma_N}. Since the frequency covariance structure of the noise in future observations cannot be estimated empirically, the forecasts assume uncorrelated Gaussian noise with $\sigma_{\rm N}=1.65\,\rm mJy\,beam^{-1}$ per channel as estimated above. The same noise model is used consistently to generate the mock observations and to construct the covariance matrix (presented in App.~\ref{app:mcovar_forecast}). In this sensitivity regime, the strongest 21-cm forest models begin to approach the noise level, as shown by the comparison with the dashed blue curve in Fig.~\ref{fig:1DPS_sim}.

\begin{figure}
    \begin{minipage}{\linewidth}
 	  \centering
 	  \includegraphics[width=\linewidth]{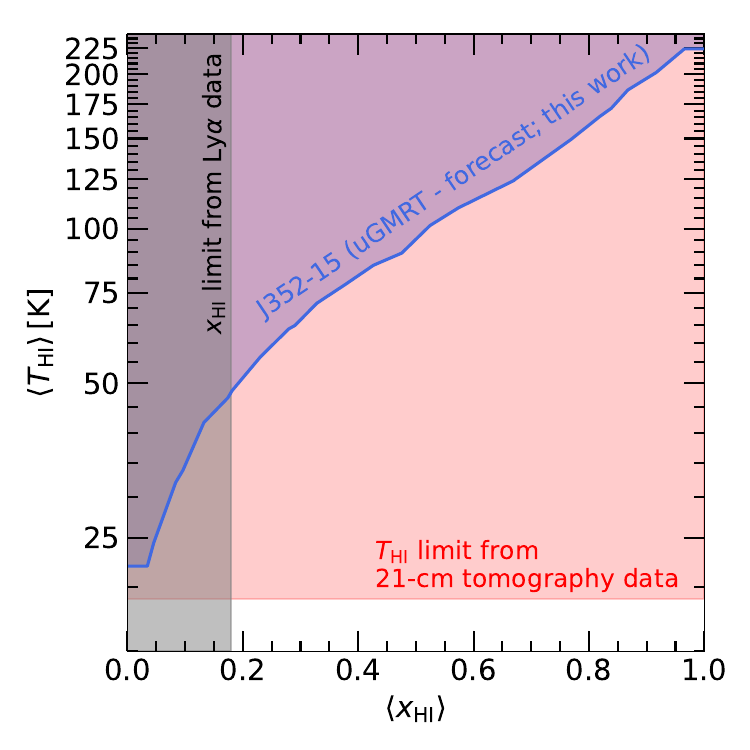}      
	\end{minipage}
    \vspace{-0.3cm}
    \caption{Similar to Fig.~\ref{fig:null-limits_archival} but based on \emph{forecasted} observations reaching a noise level of $\sigma_{\rm N}=1.65\,\mathrm{mJy\,beam^{-1}}$ achieved by a combination of archival data and forecasted new observations of $t_{\rm eff}=60\,\mathrm{hr}$.}
    \label{fig:null-limits_forecasted}
\end{figure}

On the other hand, even in the absence of a statistical detection, the increased sensitivity substantially expands the region of parameter space that can be excluded, as shown in Fig.~\ref{fig:null-limits_forecasted}. This highlights one of the key advantages of the statistical 21-cm forest framework: additional observing time remains scientifically valuable regardless of whether the signal is ultimately detected. For example, the lower limits on the neutral IGM temperature are pushed to $\langle T_{\rm HI}\rangle>[34.7,51.2,95.3,159.0]\,\rm K$ at $\langle x_{\rm HI}\rangle=[0.1,0.2,0.5,0.8]$, respectively, at the $68\%$ credible level.

These forecasts are particularly timely given that deeper Band-2 observations of J352--15 with the uGMRT are already planned (proposal ID: 50$\_$010, PI: Tomáš Šoltinský).

\section{Conclusions}\label{sec:conclusions}

In this work we present the first application of a 21-cm forest 1D power-spectrum inference framework to observational data. Using $17.5\,\rm hr$ of archival uGMRT observations of the radio-loud quasar J352--15, we calibrate the data using advanced techniques, including direction-dependent calibration, and produce a radio spectrum spanning $209.0\,\rm MHz$ to $222.5\,\rm MHz$ ($5.4\leq z\leq5.82$). From this spectrum, we obtain the first observational estimate of the 21-cm forest 1D power spectrum and use it to derive constraints on the thermal and ionization state of the largely unexplored neutral IGM near the end of the Epoch of Reionization. Our main findings can be summarized as follows:

\begin{itemize}
    \item We achieve a uGMRT Band-2 sensitivity of $\sigma_{\rm N}=3.62\,\rm mJy\, beam^{-1}$ per $6.1\,\rm kHz$ channel with effectively $11.73\,\rm hr$ on-source, post-flagging observations. This was originally $17.5\,\rm hr$, hence $\sim33\%$ of data was excised during calibration due to RFI flagging and calibration-quality filtering, including the removal of time-frequency intervals affected by residual instrumental and foreground-related systematics. 
    \item We provide the first observational estimate of the 1D power spectrum of the 21-cm forest from the spectrum of J352--15. The measured power spectrum is consistent with that derived from off-source sightlines, while a KS test of the residual-flux distributions yields $p=0.353$, indicating no statistically significant evidence for a 21-cm forest detection.  
    \item Performing Bayesian statistical analysis and assuming a null-detection of the 21-cm forest 1D power spectrum, we infer $68\%$ credible level constraints on the mean neutral fraction, $\langle x_{\rm HI}\rangle$, and the mean temperature of neutral gas,  $\langle T_{\rm HI}\rangle$. This disfavours parameter space which is allowed by the Ly$\alpha$ (forest and luminosity functions) and 21-cm (global and tomography, i.e. using CMB as radio background) line based observations. For instance, we exclude regions of $\langle x_{\rm HI}\rangle\gtrsim0.05$ for $\langle T_{\rm HI}\rangle\lesssim18\,\mathrm{K}$. At $\langle x_{\rm HI}\rangle=0.1$, a value allowed by the currently available Ly$\alpha$ observations, our 21-cm forest 1D power spectrum analysis disfavours $\langle T_{\rm HI}\rangle<26.9\,\mathrm{K}$. This is substantially stronger than the lower limits obtained from current tomographic 21-cm observations, which disfavour spin temperatures below $\sim1.3\,\mathrm{K}$ at $z=6.5$ based on MWA observations \citep{Greig_2021_MWA} and below $\sim19\,\mathrm{K}$ at $z=6$ from joint analyses incorporating multiple probes, including SARAS~3 and HERA \citep{Dhandha_2025}. At higher neutral fractions, specifically $\langle x_{\rm HI}\rangle=[0.2,0.5,0.8]$, models with $\langle T_{\rm HI}\rangle<[39.8,64.7,79.9]\,\mathrm{K}$ are excluded. These results imply that even the neutral IGM has been substantially heated above the adiabatic cooling floor.
    \item Deeper observations of J352--15 would significantly improve the scientific reach of the 21-cm forest. If the IGM remains cold and substantially neutral, a statistical detection becomes increasingly likely. Alternatively, a continued null detection would translate into substantially tighter constraints on both the thermal and ionization state of the neutral IGM, particularly disfavouring models with $\sim30-100\%$ higher $\langle T_{\rm HI}\rangle$ at $0.1\leq\langle x_{\rm HI}\rangle$.
\end{itemize}

Besides the uGMRT observations, we also analysed archival MWA observations of J352--15. These provide an independent measurement of the 21-cm forest 1D power spectrum using a different instrument, calibration strategy, and observing setup. While the substantially lower sensitivity and larger field of view of the MWA prevent it from placing competitive constraints on the thermal and ionization state of the IGM, the resulting power spectrum is consistent with noise-dominated fluctuations and demonstrates that 21-cm forest measurements can be pursued with multiple low-frequency radio facilities.

Our modelling contains four principal limitations. First, we do not account for radiative feedback from the background quasar on the surrounding IGM. For J352--15-like parameters, the affected region could extend to several tens of cMpc \citep{Soltinsky_2023}. A full treatment of proximate effects would require dedicated multifrequency radiative-transfer modelling and its impact on the 21-cm forest signal 1D power spectrum is unclear, hence this is left for future work. 
Second, we do not include unresolved small-scale absorbers such as minihalos and related dense structures. These systems may contribute additional 21-cm absorption \citep[e.g.][]{Xu_2010,Kadota_2023}, but their abundance and observable impact remain uncertain owing to photoevaporation \citep{Park_2016,Nakatani_2020,Chan_2024}, stellar-feedback processes \citep{Meiksin_2011}, and tidal or ram-pressure stripping \citep{Naruse_2024}. Accurately modelling such systems would require substantially higher spatial resolution than is currently feasible across the large cosmological volumes needed for parameter inference. 
Third, our forward modelling assumes a coeval IGM, whereas the observations probe a redshift interval of 5.4<z<5.8. Because this corresponds to the final stages of reionization, both the neutral fraction and thermal state of the IGM may evolve along the line of sight. A fully self-consistent treatment would require a lightcone-based forward-modelling framework and a corresponding extension of the inference methodology developed by \citet{Soltinsky_2025} and \citet{Patil_2026}. We therefore neglect this evolution and interpret our constraints as corresponding to an effective average IGM state across the observed redshift range.
Fourth, the ionization field in \texttt{21cmFAST} is generated using an excursion-set formalism that produces predominantly fully ionized and fully neutral cells, with only limited treatment of partially ionized structures at ionization fronts. A more realistic treatment would require simulations that combine hydrodynamics and radiative transfer, such as Sherwood-Relics \citep{Puchwein_2023} or THESAN \citep{Kannan_2022_Thesan}. However, such simulations are currently too computationally expensive to sample the large parameter space required for Bayesian inference.
Despite these limitations, excursion-set approaches such as \texttt{21cmFAST} have been extensively benchmarked against radiative-transfer simulations and shown to reproduce the large-scale morphology and statistical properties of reionization at a fraction of the computational cost \citep{Mesinger_2011_21CMFAST,Zahn_2011}. They therefore remain the most practical framework for Bayesian inference over large astrophysical parameter spaces. Exploring the impact of the above modelling assumptions on the 21-cm forest power spectrum will be an important goal of future work.


Here we also comment on the limited modelling of the systematics when forecasting future observations. We incorporate the thermal noise as a white Gaussian noise with computed $\sigma_{\rm N}$. There are other systematics which drive other spectral features. For instance, as noted in Sec.~\ref{sec:calibration}, there is a bright off-axis foreground source within the field of view of J352--15. Observations of other sources may provide cleaner fields circumventing this issue. Of the 34 currently known RLQSOs at $z>5.5$, several appear particularly promising for future 21-cm forest studies. J0309+27 as it is the second brightest known source at these redshifts ($S(\nu=147\,\rm MHz)=64.2\,\rm mJy$) and coincidentally classified as a blazar \citep{Belladitta_2020}. J0410--0139 is currently the most distant known RLQSO ($z\sim7$), providing access to a longer frequency interval through the reionization era \citep{Banados_2025}.  Finally, PSO J172+18 ($z=6.82$) is among the brightest currently known RLQSOs at $z>6.5$ \citep{Banados_2021}. Together, these sources provide attractive targets for extending 21-cm forest measurements to multiple independent sightlines and higher redshifts. Following \citet{Thyagarajan_2020}, \citet{Soltinsky_2025} and \citet{Patil_2026}, combining observations of various sources can decrease the effect of sample variance on the parameter inference too. Note that the forecast assumes that the empirical scaling given by Eq.~\ref{eq:sigma_N} remains valid when extrapolated to longer integration times. However, this scaling was derived from observations spanning a limited range of $t_{\rm eff}$, and there is no guarantee that it will continue to hold for substantially deeper observations. In particular, residual systematics, calibration errors, or persistent RFI may eventually dominate the noise budget and lead to a shallower improvement in sensitivity than assumed here. However, this can be tested only with actual observations. We also note that our flagging of strong features at the end of the data calibration pipeline might be too aggressive, potentially masking 21-cm forest absorption features. Therefore, our IGM properties constraints are conservative and further information may yet be extracted from the data. In addition, the inferred constraints are conditional on the adopted forward-modelling framework and data-processing choices, including the spectral channels masking procedure, residual bandpass subtraction, and empirical treatment of the noise covariance. A robust testing including signal injection and null tests, similar to analyses of the 3D power spectrum of the 21-cm line \citep{Mertens_2020,Trott_2020,Hera_2023}, is left for future work.

In summary, although we do not detect the 21-cm forest statistically in the spectrum of J352--15, the present data already allow us to exclude some of the coldest and most neutral IGM models within the adopted modelling framework. These constraints are complementary to those obtained from the Ly$\alpha$-based probes, which primarily constrain the highly ionized IGM and only indirectly the remaining neutral regions, and to existing 21-cm experiments, which are affected by bright Galactic foregrounds. This study demonstrates that the 21-cm forest has progressed from a largely theoretical concept to an observationally informative probe of the neutral gas. The growing sample of high-redshift radio-loud quasars, together with current and future low-frequency radio facilities, opens a promising avenue for directly constraining the thermal and ionization state of neutral IGM during the Epoch of Reionization.


\begin{acknowledgements}

The authors are grateful for the insightful discussions with Gianni Bernardi, Adélie Gorce, Martin Haehnelt, Vid Iršič, Chanasorn Kongprachaya, Leon Koopmans, Florent Mertens and Sameer Patil. The authors appreciate the work of the staff of uGMRT, which is facilitated by the National Center for Radio Astrophysics (NCRA) of the Tata Institute of Fundamental Research (TIFR), of providing the observational data (Proposals: ddtC007, ddtC219). The authors appreciate also the work of the staff of MWA, which is supported for the operation by the Australian Government (NCRIS), under a contract to Curtin University administered by Astronomy Australia Limited. TŠ acknowledges the support by the Istituto Nazionale di Astrofisica Osservatorio Astronomico di Trieste (INAF-OATs) under the Theory grant `Cosmological Investigation of the Cosmic Web' (C93C23006820005) and by the Istituto Nazionale di Fisica Nucleare (INFN) INDARK grant. GK gratefully acknowledges support from the Department of Atomic Energy, Government of India, via project RTI4012. 
This manuscript was shaped in part by the conferences/meetings including Radio Cosmology and Continuum Observations in the SKA Era: A Synergic View (code: ICTS/radiocoscon2025/04; \href{https://www.icts.res.in/program/radiocoscon2025}{https://www.icts.res.in/program/radiocoscon2025}),
The galaxy-IGM connection in the first billion years (\href{https://sites.google.com/view/galaxiesxigm}{https://sites.google.com/view/galaxiesxigm}), National Astronomy Meeting (NAM) 2025 (\href{https://conference.astro.dur.ac.uk/event/7/}{https://conference.astro.dur.ac.uk/event/7/}), Kaba Kada: Exploring the first billion years of the Universe (\href{https://www.conference-eor.com/}{https://www.conference-eor.com/}) and The Fifth National Workshop on the SKA Project - From precursors to SKAO: shaping the future of Italian radio astronomy (\href{https://indico.ict.inaf.it/event/3268/}{https://indico.ict.inaf.it/event/3268/}). This study utilized the computational resources provided by the Department of Theoretical Physics, Tata Institute of Fundamental Research (TIFR), Istituto Nazionale di Astrofisica - Osservatorio Astronomico di Trieste (INAF-OATs) and Scuola Internazionale Superiore di Studi Avanzati (SISSA). We also acknowledge the developers of publicly available software which was used in this work including \textsc{21cmFAST}  \citep{Mesinger_2011_21CMFAST,Murray_2020}, \textsc{AOFLAGGER} \citep{Offringa_2012}, \textsc{astropy} \citep{Robitaille_2013}, \textsc{CUBICAL} \citep{Kenyon_2018}, \textsc{DDFacet} \citep{Tasse_2018}, \textsc{matplotlib} \citep{Hunter_2007}, \textsc{numpy} \citep{Harris_2020}, \textsc{scipy} \citep{Virtanen_2020} and \textsc{wsclean} \citep{Offringa_2014}. During the preparation of this manuscript, the authors used OpenAI's ChatGPT to assist with language refinement and readability improvements. The authors reviewed and edited all generated text and take full responsibility for the content of the publication.

\end{acknowledgements}

\bibliographystyle{aa}
\bibliography{library}

\begin{appendix}
\nolinenumbers

\section{RFI flagging fraction across separate nights}\label{app:RFI_bynight}

\begin{figure}
    \begin{minipage}{1.\linewidth}
 	  \centering
    \includegraphics[width=\linewidth]{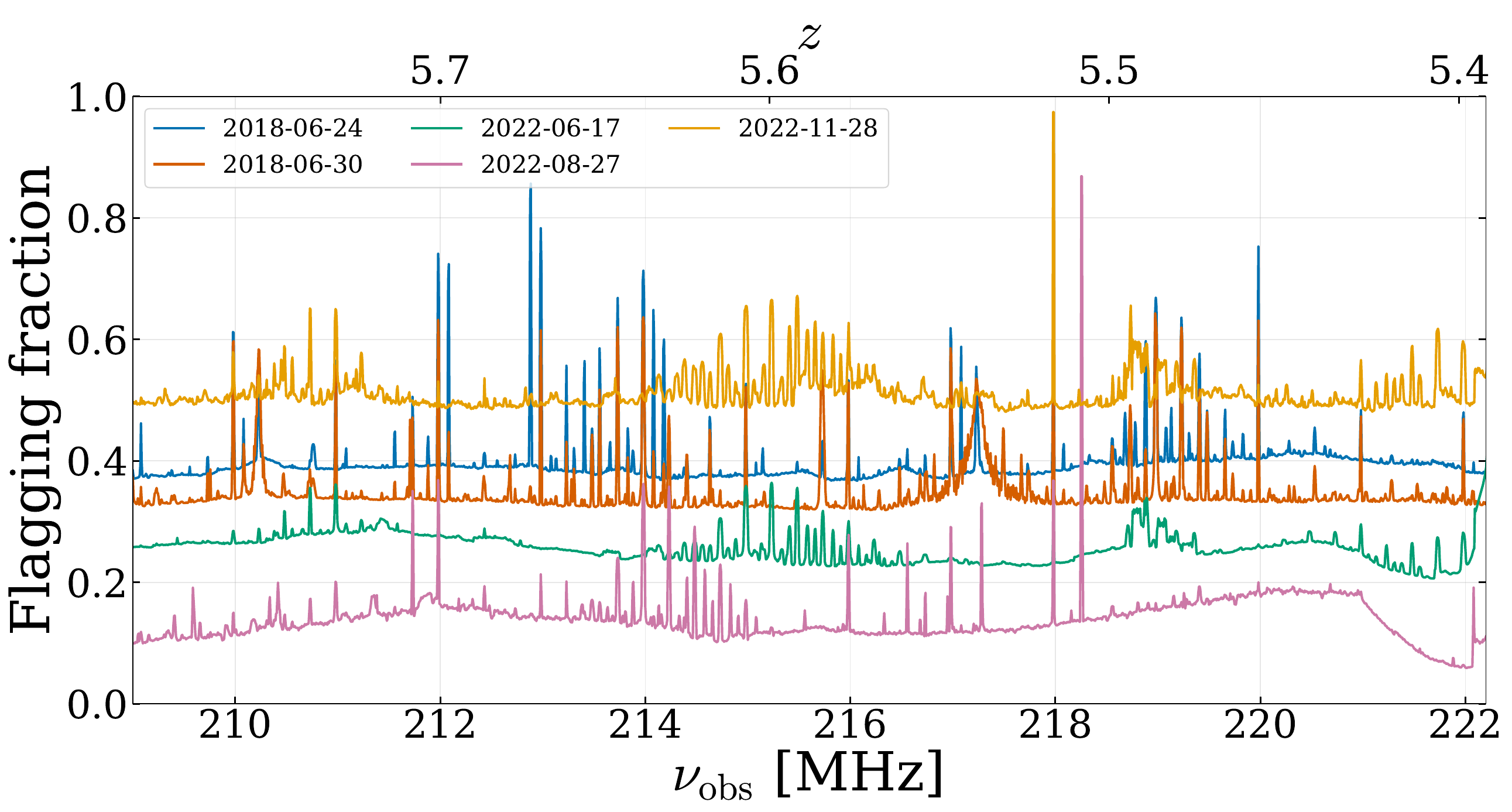}
	\end{minipage}
	\vspace{-0.3cm}
    \caption{Flagging fraction during the data excision due to RFI, self-calibration and direction-dependent effects as a function of frequency for individual observing nights.}
    \label{fig:flag_frac_bynight}
\end{figure}

The fraction of data flagged as RFI varies substantially from night to night due to changing observing conditions, including human-made interference, weather, and ionospheric activity. Figure~\ref{fig:flag_frac_bynight} shows the frequency-dependent RFI flagging fraction for each individual observing night, complementing the aggregate statistics presented in Fig.~\ref{fig:flag_frac}. The fraction of flagged data ranges from approximately $15\%$ to $50\%$ between different observing nights and can also vary significantly across frequency channels within a single observation, reflecting the varying observing conditions encountered during the campaigns. For example, the uGMRT site suffered heavy rain on the night of 28.11.2022 resulting in the reduced quality of the data.

\section{The effect of spectral resolution on the radio 1D power spectrum}\label{app:spec_res}

Increasing the frequency channel width, $\delta\nu$ only smooths out structure on smaller scales while keeping the power at larger scales unchanged. Consequently, a coarser spectral resolution reduces the maximum accessible wavenumber. Therefore, the amplitude of $\kappa_{\nu}P_{21}^{\rm S}$ is insensitive to the spectral resolution and only the accessible $\kappa_{\nu}$ range is affected.

At sufficiently high $\kappa_{\nu}$, the measured power spectrum is expected to be dominated by thermal noise (see the bottom panel of Fig.~\ref{fig:1DPS_sim}). According to the radiometer equation, the thermal-noise RMS scales as $\sigma_{\rm N}\propto\delta\nu^{-1/2}$. However, this RMS is defined per frequency channel. The corresponding noise power spectrum therefore scales as $\kappa_{\nu}P_{21}^{\rm N}\propto\sigma_{\rm N}^2\delta\nu^{-1}\propto \rm constant$. This implies that the signal-to-noise ratio is approximately independent of spectral resolution, provided that the channel width remains sufficiently small to resolve the underlying absorption features. We have verified this behaviour numerically using our forward-modelled spectra. This argument assumes that the instrumental noise is spectrally white and uncorrelated between frequency channels, and that the power spectrum estimator treats different channels as statistically independent. In practice, residual bandpass structure, calibration errors, or interpolation over flagged channels can introduce spectral correlations that violate this assumption.

The choice of $\delta\nu$ should therefore balance the formal noise properties against practical observational considerations. Sufficiently fine spectral resolution is required for effective RFI identification and excision. Conversely, excessively small channel widths produce substantially larger data volumes, increasing storage requirements and computational costs during calibration and analysis. Additional considerations include bandpass stability, spectral interpolation across flagged channels, and the potential introduction of correlated noise between neighbouring frequency bins.

\section{The effect of residual broadband features subtraction and flagging spectral channels on the radio 1D power spectrum}\label{app:flagging}

\begin{figure}
    \begin{minipage}{1.\linewidth}
 	  \centering
    \includegraphics[width=\linewidth]{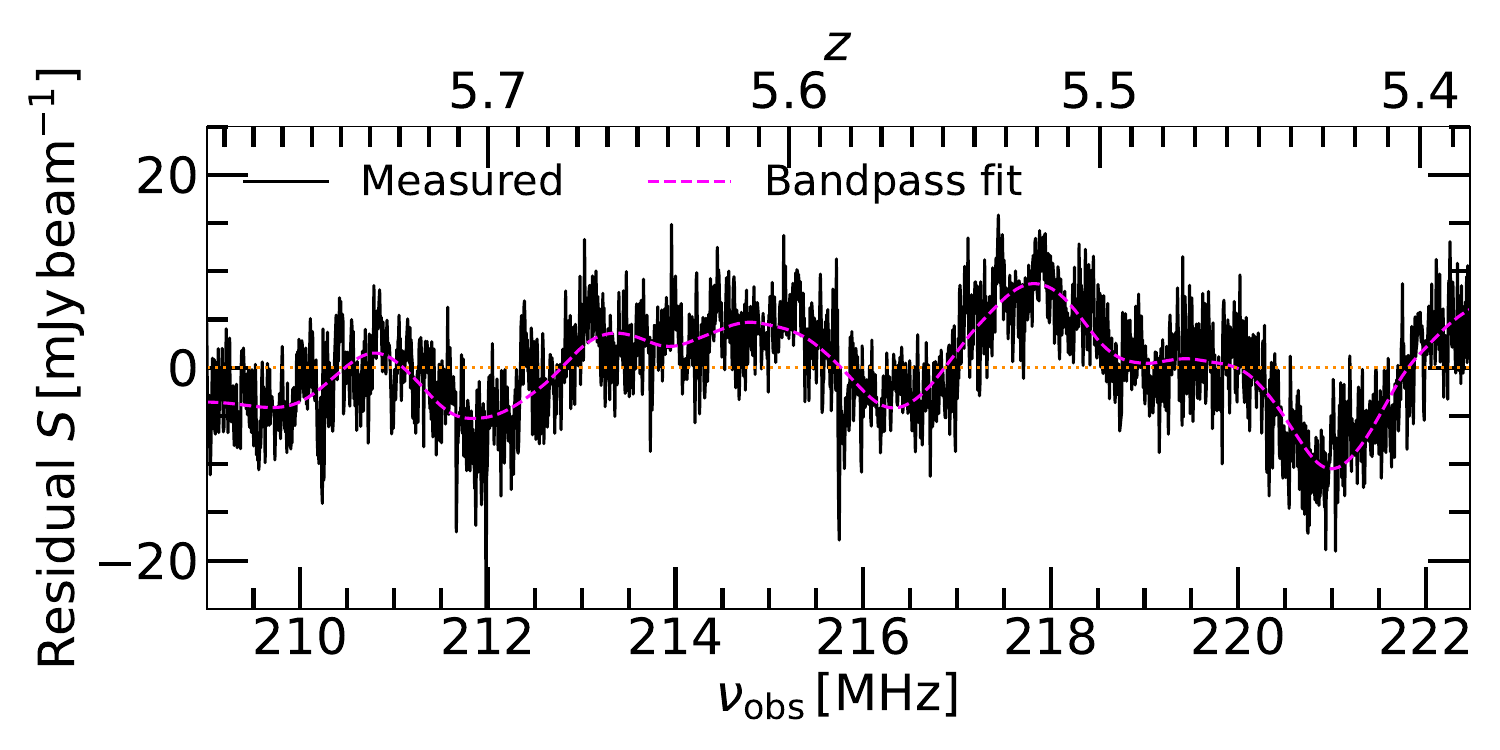} 
    \includegraphics[width=\linewidth]{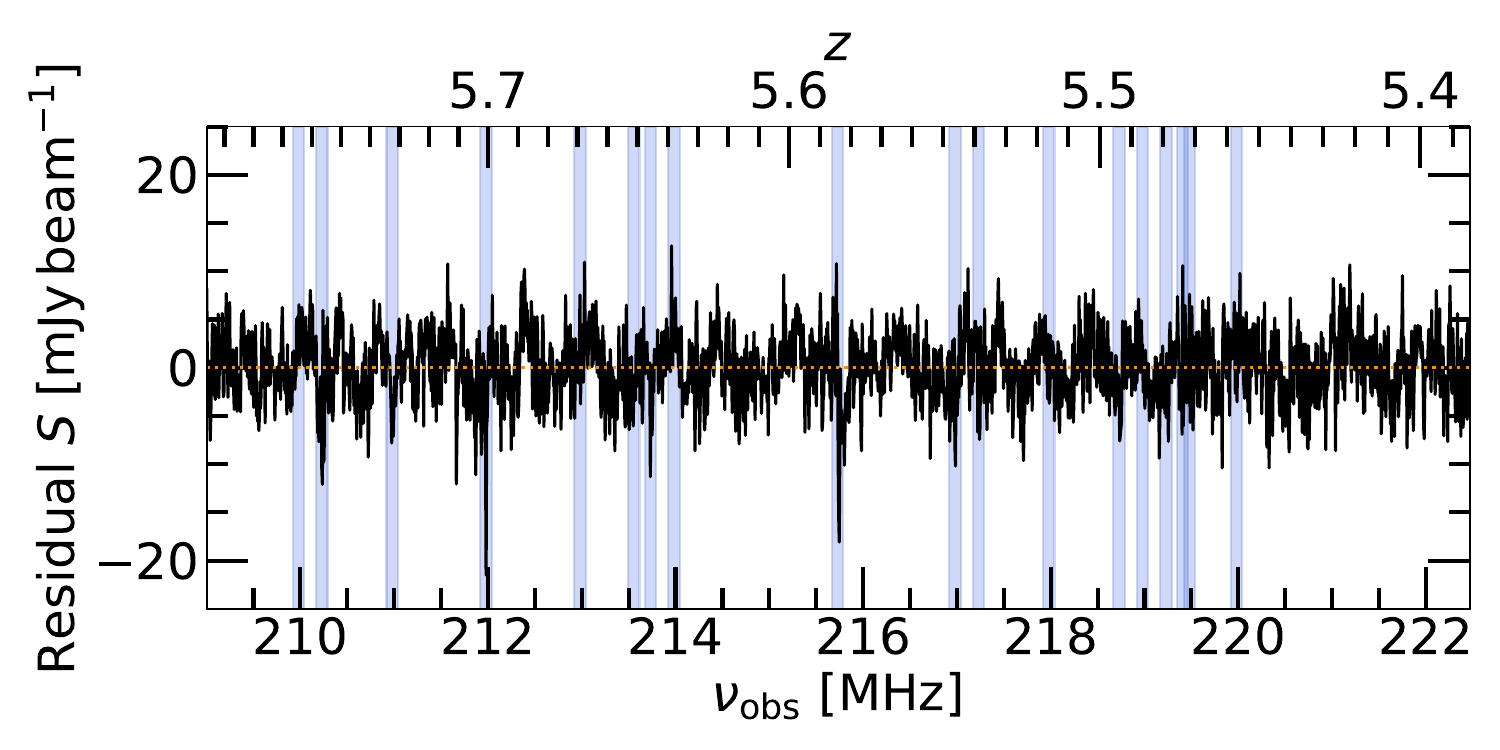}
	\end{minipage}
	\vspace{-0.3cm}
    \caption{\textit{Top:} Measured residual flux density spectrum of J352-15 (solid black curve) before the residual bandpass subtraction. The dashed pink curve indicates the fitted function to be subtracted.
    \textit{Bottom:} The same but after the residual bandpass is mitigated (solid black curve) with spectral channels flagged due significantly higher noise rms over many sightlines indicated by the blue shaded regions.}
    \label{fig:F21_flag}
\end{figure}

\begin{figure}
    \begin{minipage}{1.\linewidth}
 	  \centering
 	  \includegraphics[width=\linewidth]{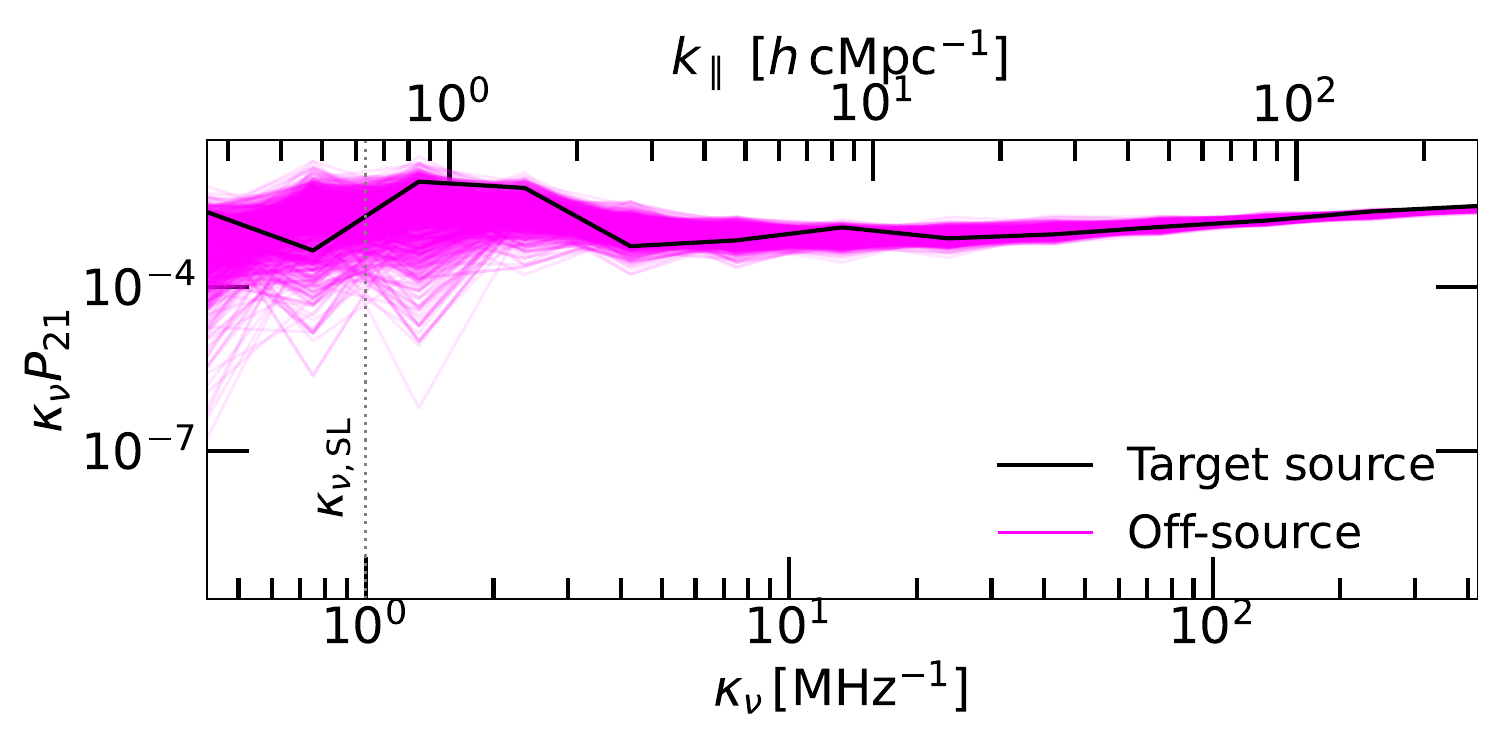}
 	  \includegraphics[width=\linewidth]{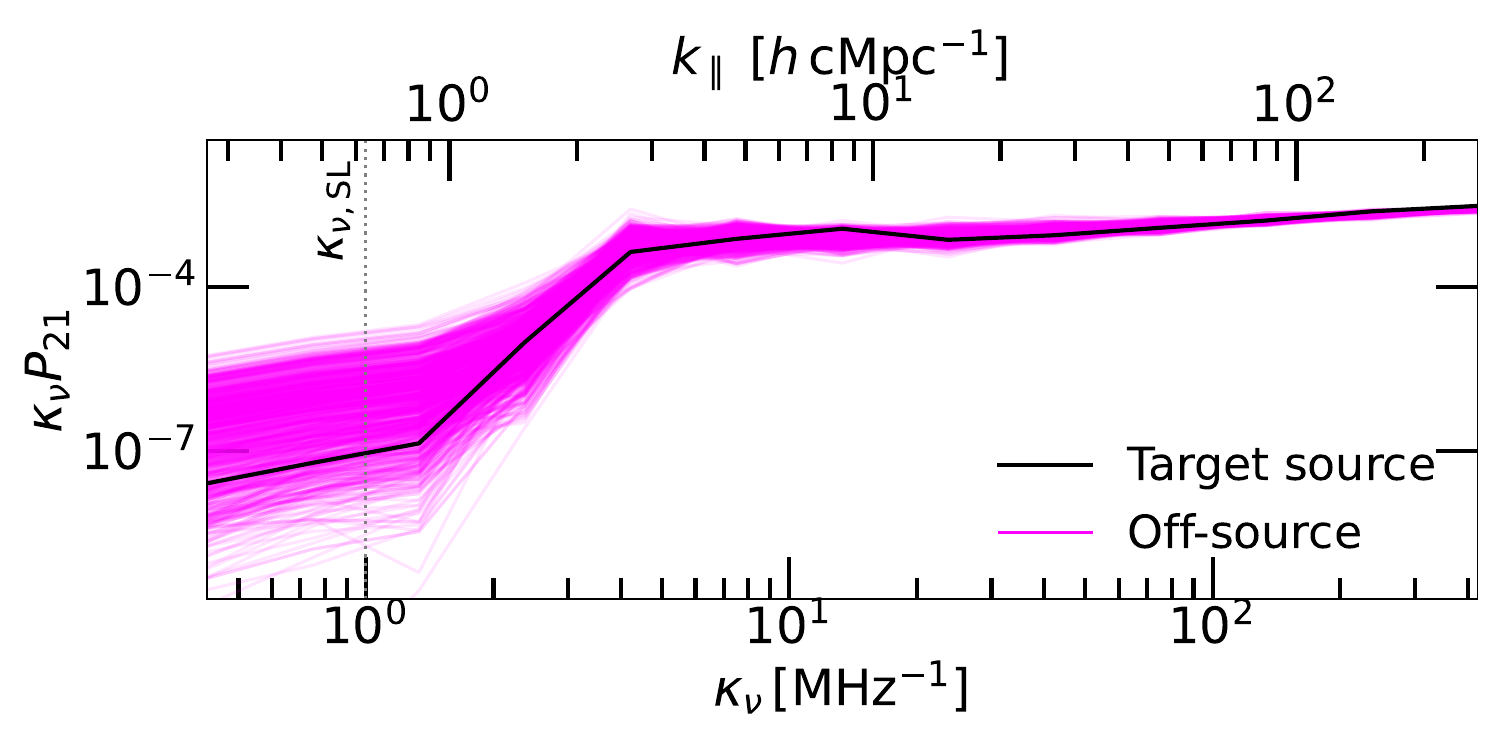}
    \includegraphics[width=\linewidth]{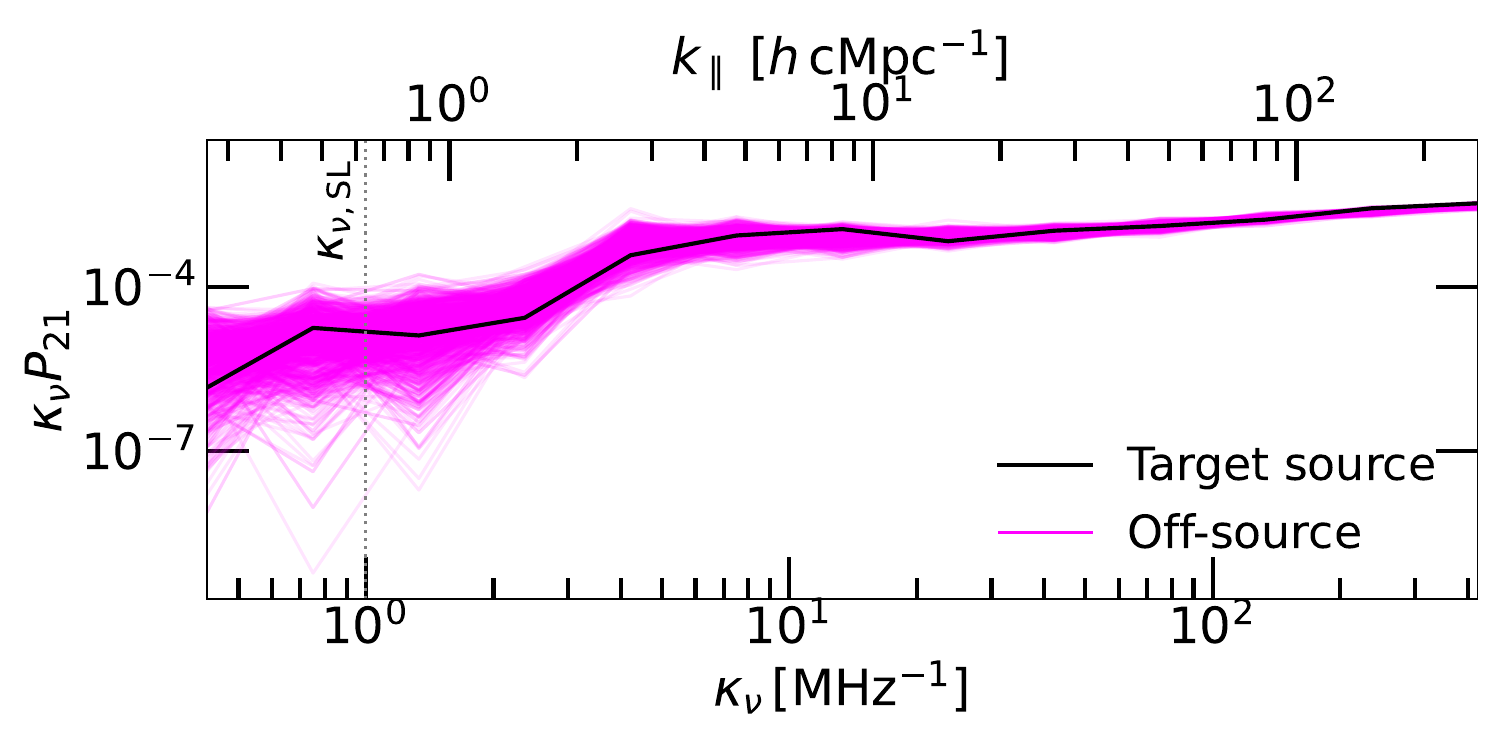}
	\end{minipage}
	\vspace{-0.3cm}
    \caption{Measured 1D power spectrum of the target (black curve) and off-source (pink curves) sightlines before residual bandpass subtraction (top panel), after it but before channels masking due to systematics (middle panel) and after applying both calibration steps (bottom panel, i.e. identical to Fig.~\ref{fig:1DPS_obs}).}
    \label{fig:1DPS_flag}
\end{figure}

The top panel of Fig.~\ref{fig:F21_flag} presents the residual flux density spectrum of J352--15 which includes the broadband features, likely arising from residual bandpass structure not fully removed by the standard calibration procedure. These features enhance power at large scale modes of $\kappa_{\nu}\lesssim3\,\rm MHz^{-1}$ as one can see by comparing the top panel of Fig.~\ref{fig:1DPS_flag} (i.e. 1D power spectrum before the residual broadband features are mitigated) with the lower two panels (i.e. 1D power spectrum after this mitigation is applied). This structure is fitted well by the coarsest level of Daubechies-8 wavelet represented by the dashed pink curve. The subtraction of this curve leads to a spectrum more suitable for a fluctuation analysis of the 21-cm forest, given that it lacks the large-scale spectral structure apparent in the original spectrum, as shown in the bottom panel of Fig.~\ref{fig:F21_flag}. Moreover, the noise RMS decreases from $5.00\,\rm mJy\,beam^{-1}$ to $3.62\,\rm mJy\,beam^{-1}$, demonstrating that residual broadband features contribute significantly to the apparent spectral variance. We report a similar result when applying Savitzky-Golay smoothing filter, but this requires a higher degree of fine-tuning. This test highlights the sensitivity of large-scale 1D power-spectrum measurements to residual broadband spectral features.

Particular frequency channels in uGMRT Band-2 exhibit significantly higher levels of noise than other ones across many measured spectra within the studied observations. As described in Sec.~\ref{sec:obs_spectrum}, channels whose frequency-dependent rms noise, $\sigma_{\rm N}(\nu)$, exceeds the threshold of $1.5\sigma_{\rm N}$ (mean noise rms across all channels) together with the surrounding 10 channels on either side are masked. The blue shaded boxes in the bottom panel of Fig.~\ref{fig:F21_flag} mark these flagged channels as shown in the measured residual flux density spectrum of J352-15. 

Here we illustrate the effect of this masking procedure on the measured 1D power spectrum in Fig.~\ref{fig:1DPS_flag}. In the top panel we show the $\kappa_{\nu}P_{21}$ computed from the residual flux spectra of target (black curve) and off-source (pink curves) sightlines without flagging these corrupted channels. The bottom panel is identical to Fig.~\ref{fig:1DPS_obs}, showing the same but after performing the masking. 

The masking primarily affects the lowest-$\kappa_{\nu}$ modes ($\kappa_{\nu}\lesssim3\,\rm MHz^{-1}$), where the measured power spectrum changes most significantly. This change is primarily driven by the modified spectral window function introduced by the masking procedure and the corresponding Lomb--Scargle power-spectrum estimation, rather than from a physical alteration of the underlying signal. At higher $\kappa_{\nu}$, the power spectra before and after masking are nearly identical. Since the lowest-$\kappa_{\nu}$ bin ($\kappa_{\nu}<\kappa_{\nu,\rm SL}=1\,\rm MHz^{-1}$) is excluded from the analysis due to sidelobe contamination, the impact of the masking procedure on the inferred IGM constraints is limited. Consequently, the adopted masking procedure trades statistical sensitivity for robustness against residual systematics. Any resulting constraints on the IGM properties should therefore be regarded as conservative.

\section{Posterior distribution from Bayesian analysis based on measured power spectrum}\label{app:corner_plot}

\begin{figure}
    \begin{minipage}{1.\linewidth}
 	  \centering
 	  \includegraphics[width=\linewidth]{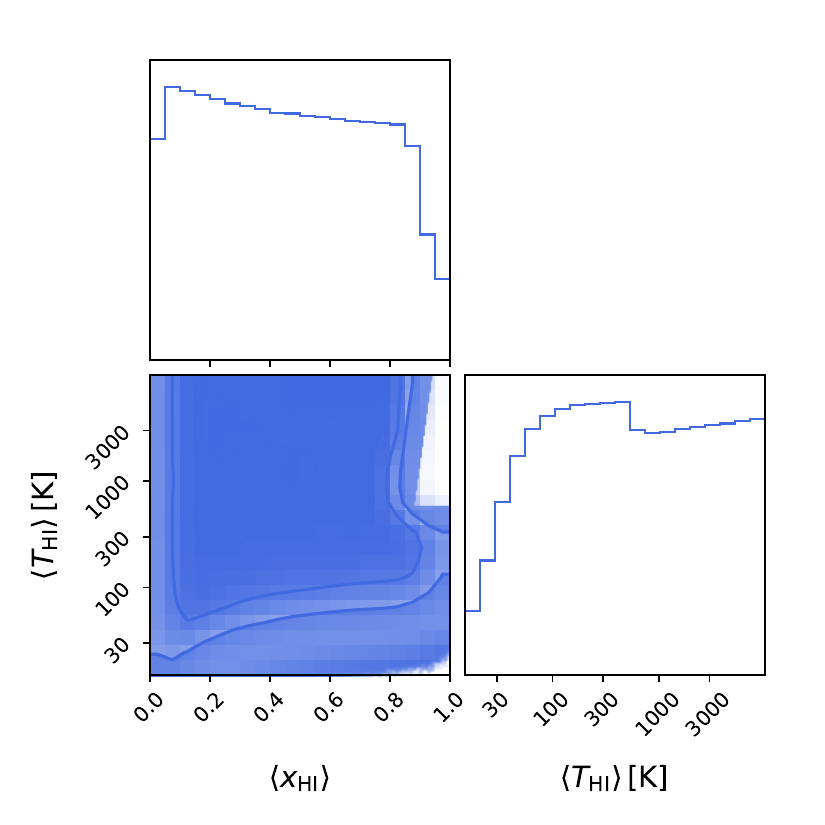}
	\end{minipage}
	\vspace{-0.3cm}
    \caption{Corner plot acquired from MCMC sampling of the posterior distribution based on our measurement of the $P_{21}$ from the archival J352--15 observations.}
    \label{fig:corner_plot_obs}
\end{figure}

We present the full corner plot resulting from our MCMC analysis based on the observed data in Fig.~\ref{fig:corner_plot_obs}. This includes the marginalized posterior distributions of $\langle x_{\rm HI}\rangle$ and $\langle T_{\rm HI}\rangle$ as well as the full extent of sampled parameter space. The top right corner of the parameter space is not sampled because our simulations do not generate unphysical models in which IGM is $\gtrsim90\%$ neutral and substantially pre-heated by the background X-ray radiation. Naturally, this is reflected in the marginalized distributions as step features at $\langle x_{\rm HI}\rangle\sim0.9$ and $\langle T_{\rm HI}\rangle\sim500\,\rm K$.

\section{Covariance matrix for forecasted observation}\label{app:mcovar_forecast}

\begin{figure}
    \begin{minipage}{1.\linewidth}
 	  \centering
 	  \includegraphics[width=\linewidth]{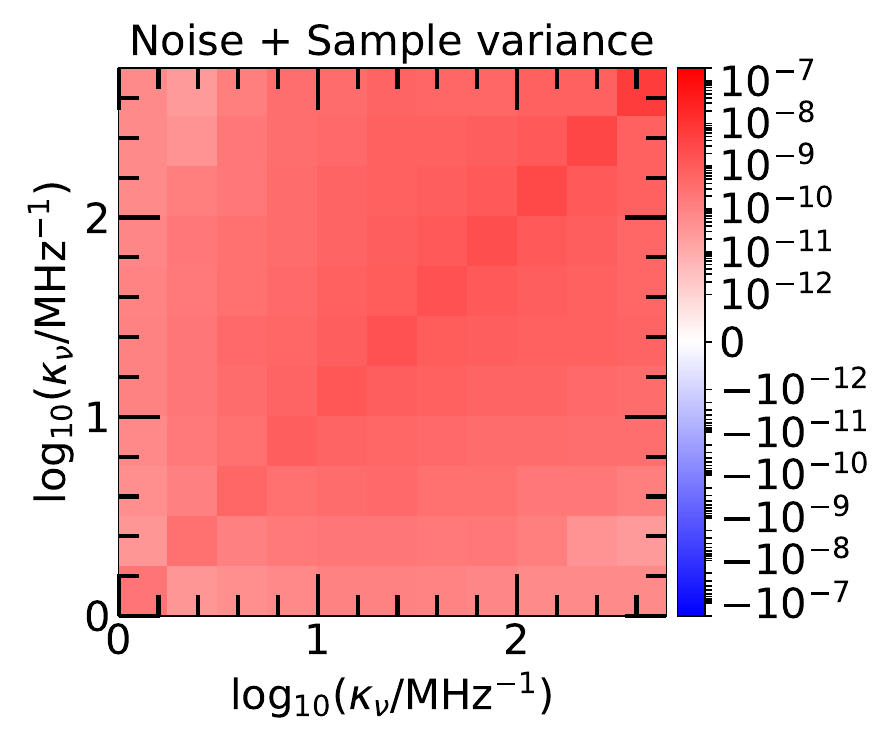}
	\end{minipage}
	\vspace{-0.3cm}
    \caption{Covariance matrix for the noise and sample variance assuming an IGM model of $\langle x_{\rm HI}\rangle=0.89$ and $\langle T_{\rm HI}\rangle=16.2\,\rm K$ and $\sigma_{\rm N}=1.65\,\rm mJy\, beam^{-1}$.}
    \label{fig:covariance_matrices_forecast}
\end{figure}

Figure~\ref{fig:covariance_matrices_forecast} shows the covariance matrix adopted for the forecast analysis described in Sec.~\ref{sec:constraints_forecasts}. The matrix is computed assuming a total effective integration time of $t_{\rm eff}=71.73\,\rm hr$, corresponding to a forecasted sensitivity of $\sigma_{\rm N}=1.65\,\rm mJy\,beam^{-1}$. It includes contributions from both instrumental noise (synthetic based on Eq.~\ref{eq:sigma_N}) and sample variance and therefore represents the forecast counterpart of the combined covariance matrix shown in the right-hand panel of Fig.~\ref{fig:covariance_matrices}.

\end{appendix}

\end{document}